\documentclass[aps,prd,superscriptaddress,nofootinbib,preprintnumbers,floatfix,notitlepage]{revtex4-1}

\usepackage{amsmath,amsthm,amssymb}
\usepackage{graphicx}
\usepackage{hyperref}
\usepackage{datetime}
\usepackage{color}
\usepackage{xcolor}
\usepackage[utf8]{inputenc}

\begin{document}

\newcommand{\amg}[1]{\textcolor{blue}{AMG: #1}}

\newcommand{\bjk}[1]{\textcolor{purple}{BJK: #1}}

\newcommand{\comm}[1]{\textcolor{green}{comm: #1}}

\title{Primordial Black Holes as a dark matter candidate}

\author{Anne M. Green}
\address{School of Physics and Astronomy, University of Nottingham,
  University Park, Nottingham, NG7 2RD, UK\\
\href{mailto:anne.green@nottingham.ac.uk}{anne.green@nottingham.ac.uk}}

\author{Bradley J. Kavanagh}
\address{Instituto de F\'isica de Cantabria (IFCA, UC-CSIC), Av.~de
Los Castros s/n, 39005 Santander, Spain\\
\href{mailto:kavanagh@ifca.unican.es}{kavanagh@ifca.unican.es}}


\begin{abstract}

The detection of gravitational waves from mergers of tens of Solar mass black hole binaries has led to a surge in interest in
Primordial Black Holes (PBHs) as a dark matter candidate. We aim to provide a (relatively) concise overview of the status of PBHs as a dark matter candidate, circa Summer 2020. First we review the formation of PBHs in the early Universe, focusing mainly on PBHs formed via the collapse of large density perturbations generated by inflation. Then we review the various current and future constraints on the present day abundance of PBHs. 
We conclude with a discussion of the key open questions in this field.
 
\end{abstract}


\maketitle

\tableofcontents

\section{Introduction}

There is convincing evidence from astronomical and cosmological observations that $\approx 85\%$ of the matter in the Universe is in the form of cold, nonbaryonic dark matter (DM), see Ref.~\cite{Bertone:2016nfn} for a historical review. The study of Primordial Black Holes (PBHs), black holes formed via the collapse of large overdensities in the early Universe, dates back to the 1960s and 70s~\cite{zn,Hawking:1971ei}. It was realised early on that PBHs are a potential DM candidate~\cite{Hawking:1971ei,Chapline:1975ojl}. 
As they form before matter-radiation equality, PBHs are non-baryonic. While PBHs are thought to evaporate via Hawking radiation~\cite{Hawking:1974sw,Hawking:1974rv}, those with initial mass $M_{\rm PBH} \gtrsim 5 \times 10^{14} \, {\rm g}$ have a lifetime longer than the age of the Universe~\cite{Page:1976df,MacGibbon:2007yq}. On cosmological scales PBH DM would behave like particle DM, however on galactic and smaller scales its granularity can have observable consequences. 

 The MACHO collaboration's 2-year Large Magellanic Cloud microlensing results~\cite{Alcock:1996yv}
  in the mid-late 1990s generated a wave of interest in PBH DM. They observed significantly more events than expected from known stellar populations. This excess was consistent with roughly half of the Milky Way (MW) halo being in 0.5 $M_{\odot}$ compact objects, with astrophysical compact objects excluded by baryon budget arguments~\cite{Fields:1999ar}. With subsequent data sets the allowed halo fraction decreased somewhat~\cite{Alcock:2000ph,Tisserand:2006zx} (see Sec.~\ref{sec:micro}). However, many of the ideas and models for producing PBH DM date back to this time.

In 2016 LIGO-Virgo announced the discovery of gravitational waves from mergers of tens of Solar mass black holes~\cite{Abbott:2016blz}.  The possibility that these BHs could be primordial rather astrophysical~\cite{Bird:2016dcv,Clesse:2016vqa,Sasaki:2016jop}, has led to a 2nd, larger, wave of interest in PBH DM. At that time Ref.~\cite{Carr:2016drx} carried out a comprehensive review of PBH DM, highlighting several mass windows where PBHs could make up all of the DM.  Subsequently there have been significant refinements of observational constraints on the abundance of PBHs. New constraints have been proposed, while some existing constraints have been weakened, or even removed completely. There have also been significant developments in
the theoretical calculations of PBH formation.

Here we aim to provide a relatively concise overview of the current (Summer 2020) status of PBHs as a dark matter candidate,~\footnote{We confine our attention to PBHs themselves as a DM candidate. The evaporation of light ($M_{\rm PBH} \lesssim 5 \times 10^{14} \, {\rm g}$) PBHs can produce stable massive particles (e.g. Ref.~\cite{Fujita:2014hha}) or leave stable Planck mass relics~\cite{MacGibbon:1987my}, both of which could also constitute the dark matter.} aimed at readers outside the field. For a comprehensive recent review of constraints on the abundances of PBHs of all masses, with an extensive reference list, see Ref.~\cite{Carr:2020gox}.  Reference~\cite{Carr:2020xqk} provides a recent overview of various potential observational consequences of PBHs, including dark matter. For a detailed review (circa 2018) of observational constraints on non-evaporated PBHs, PBH formation from large perturbations produced by inflation and PBH binaries as a source of gravitational waves see Ref.~\cite{Sasaki:2018dmp}. Reference~\cite{Khlopov:2008qy} covers formation mechanisms while Ref.~\cite{Ali-Haimoud:2019khd} focuses on future electromagnetic probes of PBHs as dark matter.

In Sec.~\ref{sec:dp}, we review the formation of PBHs, focusing mainly on the collapse of large density perturbations produced by inflation. 
In Sec.~\ref{sec:constraints}, we overview the various constraints on the present day abundance of PBHs, including potential future constraints. Finally we conclude with a summary of the current status and open questions in Sec.~\ref{sec:summary}. 
Throughout our intention is not to mention all (or even most) papers published on a topic, but to briefly describe the origin of a calculation, model or constraint and to summarise the current status. 
We set $c=1$.

\section{PBH formation}
\label{sec:dp}

The most commonly considered PBH formation mechanism is the collapse, during radiation domination, of large (adiabatic) density perturbations generated by a period of inflation
in the very early Universe. We first overview the formation of PBHs via the collapse of density perturbations during radiation domination (Sec.~\ref{sec:raddom}) and during matter domination (Sec.~\ref{sec:matdom}). We then discuss how large perturbations can be generated by inflation (Sec.~\ref{sec:large}). Finally in Sec.~\ref{sec:otherform} we briefly review other PBH formation mechanisms. We start Secs.~\ref{sec:raddom} and ~\ref{sec:large} with an overview of the essential physics, so that readers who are not interested in the subsequent details can skip them. 

\subsection{Collapse of density perturbations during radiation domination}
\label{sec:raddom}

In Sec.~\ref{sec:orig} we will first review the pioneering calculations by Carr~\cite{Carr:1975qj} of the criterion for PBH formation and the resulting PBH mass and abundance. These are sufficiently accurate for a rough understanding of the expected PBH properties. For the interested reader, we then look at refinements to the calculation of the criterion for PBH formation (Sec.~\ref{sec:deltac}), the mass of an individual PBH (Sec.~\ref{sec:mind}),
 the PBH abundance, including their mass function (Sec.~\ref{sec:betamf}) and finally spin and clustering (Sec.~\ref{sec:spincluster}). For a more extensive review of PBH formation from large inflationary perturbations, see Ref.~\cite{Sasaki:2018dmp}.
 For detailed recent studies of the relationship between the PBH abundance and the primordial power spectrum, see Refs.~\cite{Kalaja:2019uju,Gow:2020bzo}.

\subsubsection{Original calculation}
\label{sec:orig}
By considering the Jeans' length and using Newtonian gravity, Ref.~\cite{Carr:1975qj} found that a PBH will form if the density contrast $\delta \equiv \delta \rho / \rho$  in the comoving slice~\footnote{The density contrast at horizon crossing is a gauge dependent quantity. For a detailed discussion, see Sec. V of Ref.~\cite{Harada:2013epa}.}, evaluated when a given scale enters the horizon exceeds a critical, or threshold, value, $\delta_{\rm c}$, given by $\delta_{\rm c} = c_{\rm s}^2$.~\footnote{PBHs could also form on scales that never exit the horizon~\cite{Lyth:2005ze}.} Here, $c_{\rm s}$ is the sound speed, which is equal to $1/\sqrt{3}$ during radiation domination. The mass of the resulting PBH is of order the horizon mass at that time, $M_{\rm PBH} \sim M_{\rm H}$, where
\begin{equation}
M_{\rm H} \sim \frac{
t}{G} \sim 10^{15} \, {\rm g} \left( \frac{t}{10^{-23} \, {\rm s}} \right) \,.
\end{equation}
See Eq.~(\ref{mhr}) in Sec.~\ref{sec:betamf} below for a more precise expression for $M_{\rm H}$. A PBH formed at around the QCD phase transition $(t \sim 10^{-6} \, {\rm s})$ would have mass of order a Solar mass ($M_{\odot}= 2 \times 10^{30} \, {\rm g})$. These initial analytic PBH formation calculations were supported by numerical simulations a few years later~\cite{nnp}.

Reference~\cite{Carr:1975qj} also calculated the initial (i.e.~at the time of formation) abundance of PBHs
\begin{equation}
\label{beta}
\beta(M_{\rm H}) \equiv \frac{\rho_{\rm PBH}}{\rho_{\rm tot}} = \int_{\delta_{\rm c}}^{\infty} P(\delta) \, {\rm d} \delta
 \sim \sigma(M_{\rm H}) \exp{ \left( - \frac{\delta^2_{\rm c}}{2 \sigma^2(M_{\rm H})} \right)} \,,
\end{equation}
where the final step assumes that the probability distribution of primordial density perturbations, $P(\delta)$, is Gaussian with variance $\sigma^2(M_{\rm H}) \ll \delta^2_{\rm  c}$. See Sec.~\ref{sec:betamf} below for an expansion on this calculation, including a more precise definition of the mass variance, $\sigma(M_{\rm H})$, in Eq.~(\ref{massvariance}). Since the abundance of PBHs, $\beta$,  depends exponentially on the typical size of the fluctuations and the threshold for collapse, uncertainties in these quantities lead to large (potentially orders of magnitude) uncertainty in $\beta$.

\subsubsection{Criterion for PBH formation}
\label{sec:deltac}

There has been extensive work on the criterion for PBH formation in the past decade. For a more detailed review see the introduction of Ref.~\cite{Musco:2018rwt} and Sec.~2.1.~of Ref.~\cite{Shibata:1999zs}. 

Reference~\cite{Harada:2013epa} calculated the density threshold analytically finding, in the comoving slice, $\delta_{\rm c} = 0.41$ for radiation domination.
The threshold for collapse depends significantly on the shape of the density perturbation~\cite{Nakama:2013ica}. For a Gaussian density field the shape of rare peaks depends on the power spectrum~\cite{Bardeen:1985tr}. Consequently the threshold for PBH formation, and hence their abundance, depends on the form of the primordial power spectrum~\cite{Germani:2018jgr}. Recently Ref.~\cite{Musco:2018rwt} has studied a wide range of shapes and found that the threshold is lowest ($\delta_{\rm c} \approx 0.41$) for broad shapes where pressure gradients play a negligible role and highest ($\delta_{\rm c} \approx 0.67$) for peaked shapes where pressure gradients are large. The lower limit agrees well with the analytic estimate of the threshold in Ref.~\cite{Harada:2013epa}. 
Subsequently Ref.~\cite{Escriva:2019phb} has shown that the criterion for collapse is, to an excellent approximation, universal when expressed in terms of the average of the compaction function (which quantifies the gravitational potential) inside the radius at which it is maximised. If the abundance of PBHs is calculated using peaks theory (see Sec.~\ref{sec:betamf}) rather than using Eq.~(\ref{beta}) (or its more accurate form Eq.~(\ref{betaimp}) below) then the peak amplitude, rather than the average value, of the perturbation is required and this is also calculated in Ref.~\cite{Musco:2018rwt}.

Deviations from spherical symmetry could in principle affect the threshold for collapse and hence the abundance of PBHs. However the ellipticity of large peaks in a Gaussian random field is small~\cite{doroshkevich,Bardeen:1985tr}, and numerical simulations have recently shown that the effect on the threshold for collapse is negligibly small~\cite{Yoo:2020lmg}.
While non-Gaussianity of the primordial perturbations can have a significant effect on the abundance of PBHs (see Sec.~\ref{sec:betamf}), its effect on the threshold for collapse alone is also relatively small, of order a few percent~\cite{Kehagias:2019eil}.

In principle it is also possible to calculate the abundance of PBHs using the primordial curvature perturbation (which is introduced in Sec.~\ref{sec:infintro}) rather than the density contrast. However, as emphasised in Ref.~\cite{Young:2014ana} (see also Ref.~\cite{Shibata:1999zs}), perturbations on scales larger than the cosmological horizon can not (because of causality) affect whether or not a PBH forms. Consequently care must be taken if using the curvature perturbation to ensure that super-horizon modes don't lead to unphysical results. 

\subsubsection{Mass of an individual PBH}
\label{sec:mind}

It was realised in the late 1990s that, due to near critical gravitational collapse~\cite{Choptuik:1992jv}, the mass of a PBH depends on the amplitude of the fluctuation from which it forms~\cite{Niemeyer:1997mt,Niemeyer:1999ak}:
\begin{equation}
M_{\rm PBH} = \kappa M_{\rm H} (\delta - \delta_{\rm c})^{\gamma} \,,
\end{equation}
where the constants $\gamma$ and $\kappa$ depend on the shape of the perturbation and the background equation of state~\cite{Niemeyer:1999ak,Musco:2004ak}. Numerical simulations have verified that this power law scaling of the PBH mass holds down to $(\delta-\delta_{\rm c})^{\gamma}\sim 10^{-10}$~\cite{Musco:2008hv,Musco:2012au}. For PBHs formed from Mexican hat-shaped perturbations during radiation domination $\gamma=0.357$ and $\kappa = 4.02$~\cite{Musco:2008hv}. 

\subsubsection{PBH abundance and mass function}
\label{sec:betamf}

The fraction of the energy density of the Universe contained in regions overdense enough to form PBHs is usually calculated, as in Press-Schechter theory~\cite{Press:1973iz}, as~\footnote{The factor of 2 is usually included by hand to avoid the under-counting that otherwise
occurs in Press-Schechter theory.}
\begin{equation}
\beta(M_{\rm H}) = 2 \int_{\delta_{\rm c}}^{\infty} \frac{M_{\rm PBH}}{M_{\rm H}} \, P(\delta(R)) \, {\rm d} \delta(R) \,.
\end{equation}
Assuming that the probability distribution of the smoothed density contrast at horizon crossing, $\delta(R)$, is Gaussian with mass variance $\sigma(R)$ and that all PBHs form at the same time (i.e.~at the same value of $M_{\rm H}$), with the same mass $M_{\rm PBH} = \alpha M_{\rm H}$, then the PBH mass function is monochromatic and
\begin{equation}
\label{betaimp}
\beta(M_{\rm H}) = \sqrt{\frac{2}{\pi}} \frac{\alpha}{\sigma(R)} \int_{\delta_{\rm c}}^{\infty} \exp{ \left( - \frac{\delta^2(R)}{2 \sigma^2(R)}\right)} \, {\rm d} \delta(R)   = \alpha \, {\rm erfc} \left( \frac{\delta_{\rm c}}{\sqrt{2} \sigma(R)}\right) \,.
\end{equation}
The mass variance is given by~\cite{Blais:2002gw,Josan:2009qn}
\begin{equation}
\label{massvariance}
\sigma^{2}(R) = \frac{16}{81} \int_{0}^{\infty}  (k R)^4 W^2(kR)  {\cal P}_{\cal R}(k)  T^2(kR/\sqrt{3}) \, \frac{{\rm d} k}{k} \,,
\end{equation}
where $W(kR)$ is the Fourier transform of the window function used to smooth the density contrast on a comoving scale $R$, ${\cal P}_{\cal R}(k)$ is the power spectrum of the primordial comoving curvature perturbation (see e.g. Ref.~\cite{Malik:2008im}) and $T(y)$ is the transfer function which describes the evolution of the density perturbations on subhorizon scales:
\begin{equation}
T(y) = 3 \, \frac{ \left( \sin{y} - y \cos{y}\right) }{y^3} \,.
\end{equation}
The appropriate window function to use for PBH formation is not known and the relationship between the amplitude of the power spectrum and $\sigma(R)$ (and hence the abundance of PBHs formed) depends significantly on the choice of window function~\cite{Ando:2018qdb}.  For a locally scale-invariant power spectrum with amplitude ${\cal P}_{\cal R}(k) = A_{\rm PBH}$, one finds $\sigma^2(R) = b A_{\rm PBH}$ with $b =1.1, 0.09$ and $0.05$ for real-space top-hat, Gaussian and k-space top-hat window functions respectively~\cite{Ando:2018qdb}. 
The horizon mass, $M_{\rm H}$, within a comoving radius, $R$, during radiation domination is given by (c.f. Appendix A of Ref.~\cite{Wang:2019kaf})
\begin{equation}
\label{mhr}
M_{\rm H} =  5.6 \times 10^{15} M_{\odot}   \left( \frac{g_{\star, {\rm i}}}{106.75} \right)^{-1/6} (R k_{0})^2 \,.
\end{equation}
This expression has been normalised to a fiducial comoving wavenumber, $k_{0} = 0.05  \, {\rm Mpc}^{-1}$, corresponding to the Cosmic Microwave Background (CMB) pivot scale and assumes that the initial effective degrees of freedom for entropy and energy density are equal and denoted by $g_{\star, {\rm i}}$.  Excursion set (or peaks) theory~\cite{Bardeen:1985tr}, which uses the heights of peaks in the density field rather than their averaged value, can also be used to calculate the PBH abundance~\cite{Green:2004wb,Young:2014ana,MoradinezhadDizgah:2019wjf}.

In the case of critical collapse, where the mass of a PBH depends on the size of the fluctuation from which it forms (see Sec.~\ref{sec:mind}), even if all PBHs form at the same time they have an extended mass function (MF)~\cite{Niemeyer:1997mt}. While the MF is peaked close to the horizon mass, it has a significant low mass tail.

If the mass of each PBH remains constant and mergers are negligible (see Sec.~\ref{sec:gwmergers} for discussion of the latter) then the PBH density evolves with the scale factor, $a$, as $\rho_{\rm PBH} \propto a^{-3}$. During matter domination the fraction of the total density in the form of PBHs remains constant, while during radiation it grows proportional to $a$. For a monochromatic mass function the present day (at $t=t_{0}$) PBH density parameter is given by~\cite{Carr:2009jm} 
\begin{equation}
\label{omegapbh}
\Omega_{{\rm PBH}} \equiv \frac{\rho_{\rm PBH}(t_0)}{\rho_{\rm c}(t_0)} = \left( \frac{ \beta(M_{\rm H})}{1.1 \times 10^{-8}} \right)
\left( \frac{h}{0.7} \right)^{-2}  \left( \frac{g_{\star, {\rm i}}}{106.75} \right)^{-1/4} \left( \frac{M_{\rm H}}{M_{\odot}} \right)^{-1/2} \,,
\end{equation}
where $\rho_{\rm c}$ is the critical density for which the geometry of the Universe is flat, $h$ is the dimensionless Hubble constant, $H_{0} = 100\,  h \, {\rm km \, s}^{-1} \, {\rm Mpc}^{-1}$, and again it is assumed that the initial effective degrees of freedom for entropy and energy density are equal. 
Accretion of gas onto multi-Solar mass PBHs at late times can increase their mass significantly, and this needs to be taken into account when translating constraints on the present day PBH abundance into constraints on the initial PBH abundance~\cite{DeLuca:2020fpg}. Constraints arising from this accretion are discussed in Sec.~\ref{sec:accretionconstraint}.

If the primordial power spectrum has a finite width peak, then PBHs can be formed at a range of times and in this case the spread in formation times (or equivalently horizon masses at the time of formation) also needs to be taken into account. Often (e.g.~Refs.~\cite{Kuhnel:2015vtw,Clesse:2015wea,Carr:2016drx,Byrnes:2018txb,Wang:2019kaf}) the PBH mass function is calculating by binning the primordial power spectrum by horizon mass, calculating the mass function for each bin, and then summing these mass functions. The resulting mass functions can often be well fit by a lognormal distribution~\cite{Green:2016xgy,Kannike:2017bxn}. The accurate calculation of the MF resulting from a broad power spectrum is, however, an outstanding problem. In principle a region which is over-dense when smoothed on a scale $R_{1}$ could also be over-dense when smoothed on a scale $R_{2} > R_{1}$, and hence the original PBH with mass $M_{1}$ is then subsumed within a PBH of mass $M_{2}> M_{1}$. This general situation in structure formation is known as the `cloud in cloud' problem.  Reference~\cite{MoradinezhadDizgah:2019wjf} argued that for a broad power spectrum the probability that a PBH with mass $M_{1}$ is subsumed within a PBH with mass $M_{2} \gg M_{1}$ is small. For work towards an accurate calculation of the PBH MF, see e.g.~Refs.~\cite{Suyama:2019npc,Germani:2019zez}.

During phase transitions the pressure is reduced and consequently
the threshold for PBH formation, $\delta_{\rm c}$, is reduced (e.g.~Refs.~\cite{Carr:1975qj,Harada:2013epa}) and PBHs form more abundantly. In particular (if the primordial power spectrum is close to scale-invariant on small scales) the QCD phase transition leads to enhanced formation of Solar mass PBHs~\cite{Jedamzik:1996mr,Jedamzik:1999am,Byrnes:2018clq} and other phase transitions may lead to enhanced formation of PBHs with other masses~\cite{Carr:2019kxo}.

It has long been realised that since PBHs form from the high amplitude tail of the density perturbation probability distribution, non-Gaussianity can have a significant effect on their abundance~\cite{Bullock:1996at,Ivanov:1997ia}.
Reference~\cite{Franciolini:2018vbk} presents a path-integral formulation for calculating the PBH abundance (in principle) exactly in the presence of non-Gaussianity.  In practice this is non-trivial as the resulting expression depends on all of the $n$-point correlation functions. In many PBH-producing inflation models quantum diffusion is important (see Sec.~\ref{sec:large}) and in this case the high amplitude tail of the probability distribution is exponential rather than Gaussian~\cite{Pattison:2017mbe,Ezquiaga:2019ftu}.
It should be noted that even if the underlying curvature perturbations are Gaussian, the non-linear relationship between density and curvature perturbations inevitably renders the distribution of large density perturbations non-Gaussian~\cite{Kawasaki:2019mbl,DeLuca:2019qsy,Young:2019yug}.

\subsubsection{Spin and clustering}
\label{sec:spincluster}

The rare high peaks in the density field from which PBHs form are close to spherically symmetric. Therefore the torques on the collapsing perturbation, and the resulting angular momentum, are small. Consequently PBHs are formed with dimensionless spin parameters, $a = |{\bf S}|/(G M_{\rm PBH}^2)$ where ${\bf S}$ is the spin, of order $0.01$ or smaller~\cite{Mirbabayi:2019uph,DeLuca:2019buf}. Note however that accretion of gas at late times may increase the spin of massive ($M_{\rm PBH} \gtrsim 30 M_{\odot}$) PBHs~\cite{DeLuca:2020bjf}.

As PBHs are discrete objects there are Poissonian fluctuations in their distribution~\cite{Afshordi:2003zb}.
The initial clustering of PBHs was first studied in Refs.~\cite{Afshordi:2003zb,Chisholm:2005vm}, which found that PBHs would be formed in clusters. More recently it has been shown that if the primordial curvature perturbations are Gaussian and have a narrow peak then the PBHs are not initially clustered, beyond Poisson~\cite{Ali-Haimoud:2018dau,Desjacques:2018wuu,Ballesteros:2018swv}.
Reference~\cite{MoradinezhadDizgah:2019wjf} has argued that the initial clustering is also small for broad spectra. However local non-Gaussianity~\footnote{In local non-Gaussianity the probability distribution of the primordial fluctuations is a local function of one or more Gaussian random fields. Non-negligible local non-Gaussianity can arise if there are multiple light scalar fields present during inflation, e.g. Ref.~\cite{Wands:2010af}.}  can lead to enhanced initial clustering
~\cite{Tada:2015noa,Young:2015kda,Suyama:2019cst}. 

\subsection{Collapse of density perturbations during matter domination}
\label{sec:matdom}

It is usually assumed that the evolution of the  Universe is radiation dominated from the end of inflation up until matter-radiation equality at $t_{\rm eq} = 1.7 \times 10^{12} \, {\rm s}$. However it is possible that there could be a period of matter domination (with equation of state $p \approx 0$) prior to Big Bang Nucleosynthesis due to, for instance, long-lived particles dominating the Universe and then decaying (see e.g.~Ref.~\cite{Khlopov:1980mg,
Georg:2016yxa,Carr:2017edp,Allahverdi:2020bys} for discussion in the context of PBH formation).

The criteria for PBH formation during matter domination are significantly different than during radiation domination. During matter domination the density contrast grows as $\delta \propto a$, so that in principle small perturbations can grow sufficiently to form a PBH. However a perturbation has to be (close) to spherical and homogeneous for a PBH to form~\cite{Khlopov:1980mg}. The modified expansion history of the Universe also modifies the relationship between the initial PBH mass fraction, $\beta$, and the present day PBH density parameter, $\Omega_{\rm PBH}$, Eq.~(\ref{omegapbh}).

The fraction of horizon-sized regions which  collapse to form a PBH can be written as the product of the fraction of regions which separately satisfy the inhomogeneity and anisotropy criteria: $\beta = \beta_{\rm inhom}  \times \beta_{\rm aniso}$~\cite{Kokubu:2018fxy}. If a perturbation is not sufficiently spherically symmetric it will collapse to form a pancake or cigar. Khlopov and Polnarev originally found $\beta_{\rm aniso} \approx 0.02 \sigma^{5}$, where $\sigma$ is the mass variance as defined in Eq.~(\ref{massvariance}). Reference~\cite{Harada:2016mhb}  revisited this calculation numerically. Their result, which is valid for all $\sigma$, can be approximated by $\beta_{\rm aniso} \approx 0.056 \sigma^{5}$ for $\sigma \lesssim 0.01$. 
Polnarev and Khlopov argued that for a PBH to form, a fluctuation must collapse to within its Schwarzschild radius before a caustic can form at its centre, and found that the fraction of regions which satisfy this criteria is given by $\beta_{\rm inhom} \approx \sigma^{3/2}$~\cite{Khlopov:1981}. Taking into account the finite propagation speed of information $\beta_{\rm inhom} \approx 3.7 \sigma^{3/2}$ (for $\sigma \ll 1$)~\cite{Kokubu:2018fxy}. The final result (for $\sigma \ll 1$) is $\beta \approx 0.21 \sigma^{13/2}$~\cite{Kokubu:2018fxy}, and if $\sigma \lesssim 0.05$, PBHs form more abundantly during matter domination than during radiation domination.

As angular momentum plays a significant role in PBH formation during matter domination, they are formed with large spins: $a \gtrsim 0.5$, with the exact value depending on the duration of the period of matter domination and also $\sigma$~\cite{Harada:2017fjm}. Since PBHs formed during matter domination don't form from the high peaks of the density field, local primordial non-Gaussianity would lead to smaller initial clustering than for formation during radiation domination, however it can still be much larger than the Poisson shot noise~\cite{Matsubara:2019qzv}.

\subsection{Generation of large primordial perturbations by inflation}
\label{sec:large}

Inflation is a period of accelerated expansion ($\ddot{a}> 0$, where $\dot{}$ denotes a derivative with respect to time)  in the early Universe, originally proposed to solve various problems with the standard Big Bang. It also provides a mechanism for generating primordial perturbations, via quantum fluctuations of scalar fields. 
For a comprehensive and comprehensible introduction to inflation see Baumann's lecture notes~\cite{Baumann:2009ds}.
In Sec.~\ref{sec:infintro} we review the aspects of inflation that are relevant to PBH formation and briefly discuss the requirements for generating large, PBH-producing perturbations. For the interested reader, we then look at PBH producing single (Sec.~\ref{sec:single}) and multi (Sec.~\ref{sec:multi}) field inflation models in more detail. A significant number of PBH-producing inflation models have been proposed. We will not attempt a detailed study of all possible models, but instead focus on examples of the types of models that can produce large primordial perturbations from a phenomenological point of view. In many cases the initial ideas (a plateau in the potential of a single field~\cite{Ivanov:1994pa}, hybrid inflation~\cite{Randall:1995dj,GarciaBellido:1996qt}, double inflation~\cite{Silk:1986vc,Kawasaki:1997ju} and a spectator field~\cite{Yokoyama:1995ex}) were proposed in the 1990s, motivated by the
excess of LMC microlensing events observed by the MACHO collaboration~\cite{Alcock:1996yv}. In recent years, motivated by the LIGO-Virgo discovery of massive BH binaries, these models have been revisited and refined, taking into account theoretical and observational developments in the intervening decades.

\subsubsection{Introduction}
\label{sec:infintro}
In most models of inflation the accelerated expansion is driven by a scalar field known as the inflaton. The Friedmann equation for the expansion of a universe dominated by a scalar field $\phi$ with potential $V(\phi)$ is
\begin{equation}
\label{Friedmann}
H^2 = \frac{8 \pi}{3 m_{\rm pl}^2} \left[ \frac{1}{2} \dot{\phi}^2 + V(\phi) \right] \,,
\end{equation}
and the evolution of the scalar field is governed by the Klein-Gordon equation
\begin{equation}
\label{kg}
\ddot{\phi} + 3 H \dot{\phi} + \frac{{\rm d} V}{{\rm d} \phi} =0  \,.
\end{equation}
The dynamics of inflation are often studied using the (potential) slow-roll parameters $\epsilon_{V}$ and $\eta_{V}$~\footnote{The Hubble slow-roll parameters, defined in terms of the Hubble parameter and its derivatives, allow for a more accurate calculation of the power spectrum and also a more accurate definition of the condition for accelerated expansion (see e.g.~Ref.~\cite{Baumann:2009ds}). We use the potential slow-roll parameters here because their definition in terms of the potential is initially more intuitive.} 
\begin{equation}
\epsilon_{V} \equiv \frac{m_{\rm Pl}^2}{16 \pi} \left( \frac{V^{\prime}}{V} \right)^2 \,, \hspace{2.0cm}
  \eta_{V} \equiv \frac{m_{\rm Pl}^2}{8 \pi} \left( \frac{V^{\prime \prime}}{V} \right) \,,
\end{equation}
where ${}^{\prime}$ denotes derivatives with respect to $\phi$. Accelerated expansion occurs when $\epsilon_{V} \lesssim 1 $. In the slow-roll approximation (SRA), $\epsilon_{V}$ and $ \eta_{V}$ are both much less than one, and the $\dot{\phi}$ term in the Friedmann equation (Eq.~(\ref{Friedmann})) and the $\ddot{\phi}$ term in the Klein-Gordon equation (Eq.~(\ref{kg})) are negligible. In this regime the power spectrum of the primordial curvature perturbation, ${\cal P}_{\cal R}(k) \equiv (k^3/2 \pi^2) \langle |{\cal R}_{k}| \rangle$, is given by
\begin{equation}
\label{psv}
{\cal P}_{\cal R}(k) \approx \frac{ 8 }{3 m_{\rm pl}^4} \frac{V}{\epsilon_{V}} \,,
\end{equation}
where $V$ and $\epsilon_{V}$ are to be evaluated when the scale of interest exits the horizon during inflation, $k=aH$. It is common to use a Taylor
expansion of the spectral index to parameterise the primordial power spectrum on cosmological scales ($k_{\rm cos} \approx (10^{-3}-1) \, {\rm Mpc}^{-1}$):~\footnote{N.b.~this approach should not be used to extrapolate down to PBH DM forming scales, $k_{\rm PBH} \sim (10^{5}-10^{15}) \, {\rm Mpc}^{-1}$, as the expansion does not converge over this much wider range of scales~\cite{Green:2018akb}.}
\begin{equation}
{\cal P}_{\cal R}(k) = A_{\rm s} \left( \frac{k}{k_{0}} \right)^{(n_{\rm s}(k) -1)} \,,
\end{equation}
where $k_{0}$ is the pivot scale about which the expansion is carried out and
\begin{equation}
n_{\rm s}(k)= n_{\rm s}|_{k_{0}} + \frac{1}{2} \left. \frac{{\rm d} n_{\rm s}}{{\rm d} \ln{k}} 
\right|_{k_{0}} \ln{\left( \frac{k}{k_{0}} \right)} + ... \,. 
\end{equation}
In the SRA, $n_{\rm s} |_{k_{0}} -1 = 2 \eta_{V} - 6\epsilon_{V}$. Primordial tensor modes, which manifest as gravitational waves, are also generated by inflation and in the SRA the ratio of the amplitudes of the tensor and scalar power spectra on cosmological scales (known as the `tensor to scalar ratio') is given by $r \equiv {\cal P}_{\rm t}(k_{0})/{\cal P}_{\cal R}(k_{0}) = 16 \epsilon_V$.

The amplitude and scale dependence of the primordial perturbations on cosmological scales are now accurately measured. In particular a scale-invariant power spectrum ($n_{\rm s}=1$) is excluded at high confidence. From Planck 2018, combined with other CMB and large scale structure data sets~\cite{Akrami:2018odb}:
\begin{eqnarray}
\ln{(10^{10} A_{\rm s})} &=& 3.044 \pm 0.0014 \,,  \\
n_{\rm s} |_{0.05 \, {\rm Mpc}^{-1}}  &=& 0.9668 \pm 0.0037 \,, \\
r &<& 0.063 \,,
\end{eqnarray}
where $1\sigma$ errors on measured parameters and a $95\%$ upper confidence limit on $r$ are stated.
The upper limit on $r$ leads to a relatively tight constraint on the slope of the potential in the region that corresponds to cosmological scales, $\epsilon_{V} < 0.0039$, and various inflation models are tightly constrained, or excluded (e.g. $V(\phi) \propto \phi^2$)~\cite{Akrami:2018odb}.
There are also constraints on smaller scales from spectral distortions of the CMB and induced gravitational waves (see Sec.~\ref{sec:indirect} for a more detailed discussion). However the current constraints on the amplitude of the power spectrum on small scales are fairly weak. The COBE/FIRAS limits on spectral distortions require ${\cal P}(k) \lesssim 10^{-4}$~\cite{Fixsen:1996nj,1994ApJ...420..439M} for $k \approx (10-10^{4}) \, {\rm Mpc}^{-1}$ and Pulsar Timing Array (PTA) limits on gravitational waves require ${\cal P}(k) \lesssim 10^{-2}$ for $k \approx (10^{6}-10^{7}) \, {\rm Mpc}^{-1}$~\cite{Byrnes:2018txb,Inomata:2018epa,Chen:2019xse}.
However a future PIXIE-like experiment could tighten the spectral distortion constraint to ${\cal P}(k) \lesssim 10^{-8}$~\cite{Chluba:2019nxa} and SKA and LISA will improve the current induced gravitational wave constraints over a wide range of smaller scales~\cite{Byrnes:2018txb,Inomata:2018epa}.

On cosmological scales the amplitude of the primordial curvature perturbation power spectrum has 
been measured to be $A_{\rm s} =2.1 \times 10^{-9}$~\cite{Akrami:2018odb}. If the power spectrum were completely scale invariant~\footnote{As we saw above in fact on cosmological scales the power spectrum is `red', $n_{\rm s} < 1$, with amplitude decreasing with increasing wavenumber $k$.} then using Eq.~(\ref{betaimp}), and assuming $\delta_{\rm c}=0.5$ and $\sigma^2 = A_{\rm s}$, which is sufficient for a rough calculation, the initial mass fraction of PBHs formed during radiation domination would be $\beta \approx {\rm erfc}(7000) \approx \exp{[(-7000)^2]}/7000$, i.e.~completely negligible.
Conversely if all of the DM is in PBHs with $M_{\rm PBH} \sim M_{\odot}$ then, from Eq.~(\ref{omegapbh}), the initial PBH mass fraction must be $\beta \sim 10^{-8}$. From Eq.~(\ref{betaimp}) this requires $\delta_{c}/(\sqrt{2} \sigma) \sim 4$, and hence the mass variance on the corresponding scale must be $\sigma \sim 0.1$. This requires the amplitude of the primordial power spectrum on this scale to be $A_{\rm PBH} \sim 0.01$, 7 orders of magnitude larger than its measured value on cosmological scales. We note here that since $\beta$ depends exponentially on the amplitude of the perturbations, fine-tuning is required to achieve an interesting (i.e.~neither negligible nor unphysically large, $\Omega_{\rm PBH} \gg 1$) abundance of PBHs. Furthermore to produce PBHs of a particular mass the peak in the power spectrum must occur on a specific scale, given by Eq.~(\ref{mhr}).

\begin{figure}[t]
\begin{center}
\includegraphics[width=1.0\textwidth]{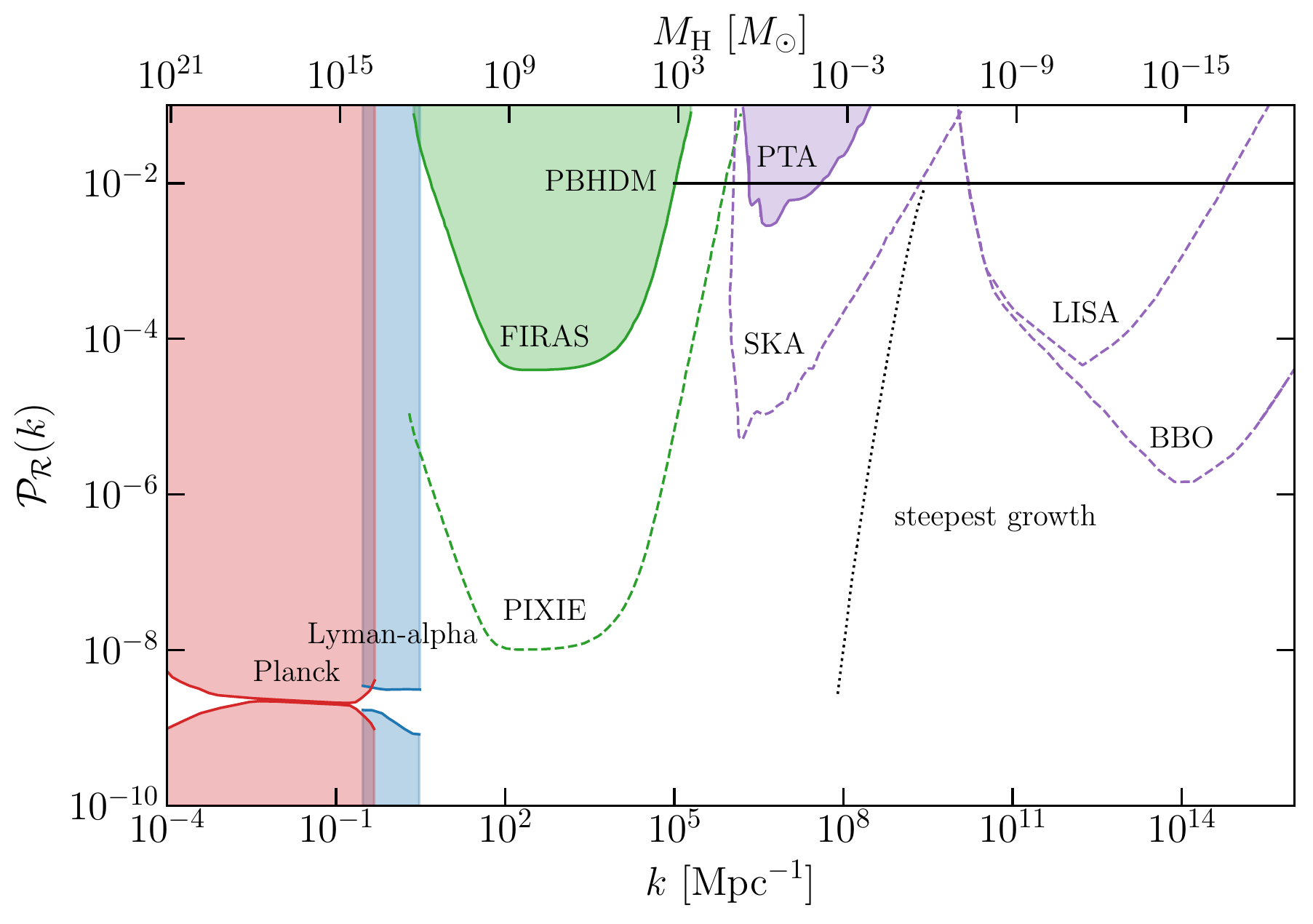}\\
\end{center}
\caption{Constraints on the primordial power spectrum ${\cal P}_{\cal R}(k)$ from the CMB temperature angular power spectrum~\cite{Akrami:2018odb} (red line), from the Lyman-$\alpha$ forest~\cite{Bird:2010mp} (blue), CMB spectral distortions~\cite{Fixsen:1996nj,1994ApJ...420..439M} (green) and pulsar timing array limits on gravitational waves~\cite{Byrnes:2018txb} (magenta). Potential constraints from a future PIXIE-like spectral distortion experiment~\cite{Chluba:2019nxa} (blue) and limits on gravitational waves from SKA, LISA and BBO (magenta)~\cite{Inomata:2018epa,Chluba:2019nxa} are shown as dotted lines. In each case the excluded regions are shaded. The spectral distortion and induced gravitational wave constraints depend on the shape of the primordial power spectrum, a $k^4$ growth followed by a sharp cut-off has been assumed here~\cite{Byrnes:2018txb}. The approximate amplitude, ${\cal P}_{\cal R}(k) \sim 10^{-2}$, required to form an interesting number of PBHs is shown as a black line (see text for details). The dotted black line shows the steepest growth possible in a single field inflation model~\cite{Byrnes:2018txb,Carrilho:2019oqg} (see Sec.~\ref{sec:single}). Adapted from Refs.~\cite{Byrnes:2018txb,Inomata:2018epa,Chluba:2019nxa}. }
\label{fig:Pk}
\end{figure}

Figure~\ref{fig:Pk} compares the amplitude of the power spectrum required on small scales to form PBH DM (${\cal P}_{\cal R}(k) \sim 10^{-2}$)~\footnote{The required amplitude is in fact scale dependent. Lighter PBHs form earlier and therefore their density relative to the total density grows for longer. Consequently the initial PBH density corresponding to a given present day density is smaller, and hence the amplitude of the power spectrum required for PBHs to make up all of the DM is smaller. However this scale dependence is smaller than the uncertainties discussed in Sec.~\ref{sec:raddom}, and therefore we do not include it here.} with the current measurements on cosmological scales from the CMB temperature angular power spectrum~\cite{Akrami:2018odb} and the Lyman--$\alpha$ forest~\cite{Bird:2010mp}, and current and future constraints on smaller scales~\cite{Byrnes:2018txb,Inomata:2018epa,Chluba:2019nxa}, which were mentioned above.~\footnote{To translate from $k$ to $M_{\rm H}$ we use Eq.~(\ref{mhr}), from Ref.~\cite{Wang:2019kaf}, which agrees with the conversion given in Ref.~\cite{Kalaja:2019uju}. However we note a discrepancy with Ref.~\cite{Byrnes:2018txb}.} 
The steepest growth of the power spectrum which can be achieved in single field inflation~\cite{Byrnes:2018txb,Carrilho:2019oqg} (see Sec.~\ref{sec:single} for discussion) is also shown. As discussed in more detail in Sec.~\ref{sec:indirect}, the current constraints already indirectly exclude $M_{\rm PBH}/M_{\odot} \gtrsim 10^{3} $ and $ 10^{-2} \lesssim M_{\rm PBH}/M_{\odot} \lesssim 1 $ for PBH DM formed from the collapse of large inflationary density perturbations during radiation domination. Significant improvements in these indirect probes in the future will cover most of the mass range for PBHs formed via this mechanism.

As we saw in Eq.~(\ref{psv}), in the SRA the power spectrum is inversely proportional to $\epsilon_{V}$ and hence the slope of the potential, so that reducing the slope increases the amplitude of the perturbations. The potential then has to steepen again so that 
$\epsilon_{V}$ becomes greater than 1 and inflation ends. This can be achieved by inserting a plateau or inflection point feature in the potential (see Sec.~\ref{sec:single}). 
Alternatively large perturbations can be generated in multi-field models.
In this case typically a different field is responsible for the perturbations on small scales than on cosmological scales. This effectively decouples the constraints on cosmological scales from the requirements for generating large perturbations.

\subsubsection{Single-field models}
\label{sec:single}

In this subsection we discuss various ways of generating large PBH forming perturbations in single-field inflation models, namely features in the potential, running of the inflaton mass, hilltop models, and the reheating period at the end of inflation.

The possibility of producing PBHs by inserting a plateau in the potential was explored in the 1990s by Ivanov et al.~\cite{Ivanov:1994pa}. More recently Ref.~\cite{Garcia-Bellido:2017mdw} showed that a PBH-forming peak in the power spectrum 
could be produced with a potential possessing an inflection point. Such a potential can be generated from a ratio of polynomials~\cite{Garcia-Bellido:2017mdw} or for a non-minimally coupled scalar field with a quartic potential~\cite{Ballesteros:2017fsr}. 
Reference~\cite{Hertzberg:2017dkh} showed that if a quintic potential is fine-tuned so that a local minimum is followed by a small maximum (so that the field is slowed down but can still exit this region) a sufficiently large peak in the power spectrum can be produced.

As shown in Ref.~\cite{Motohashi:2017kbs}, for the power spectrum of canonical single-field models to grow by the required seven orders of magnitude the SRA has to be violated. 
In the limit where the potential is flat a phase of non-attractor ultra-slow roll (USR) inflation occurs~\cite{Tsamis:2003px,Kinney:2005vj}, where the evolution of the inflaton (Eq.~(\ref{kg})) is governed by the expansion rate, rather than the slope of the potential. In this case the standard calculation of the power spectrum is not valid. In particular the curvature perturbations grow rapidly on superhorizon scales, rather than remaining constant, or `frozen out'. As emphasised by e.g.~Refs.~\cite{Germani:2017bcs,Ballesteros:2017fsr,Hertzberg:2017dkh}, in models with an inflection point, or shallow local minimum, a numerical calculation using the Sasaki-Mukhanov equations~\cite{Sasaki:1986hm,Mukhanov:1988jd} is required to accurately calculate the position and height of the resulting peak in the power spectrum. Also in this regime quantum diffusion (where quantum kicks of the inflaton field are larger than the classical evolution) occurs and has a significant effect on the probability distribution of the primordial perturbations, and hence the number of PBHs formed~\cite{Ivanov:1997ia,Pattison:2017mbe,Biagetti:2018pjj,Ezquiaga:2019ftu}.

Reference~\cite{Byrnes:2018txb} studied the fastest possible growth of the primordial power spectrum that can be achieved in principle in single-field inflation with a period of USR, finding ${\cal P}(k) \propto k^{4}$. Reference~\cite{Carrilho:2019oqg} subsequently showed that ${\cal P}(k) \propto k^{5} (\log{k})^2$ can be achieved for a specific form of the pre-USR inflationary expansion. 
These constraints on the growth of the power spectrum are important because to form an interesting abundance of PBHs the amplitude has to grow significantly from its value on cosmological scales, while evading the constraints from spectral distortions on intermediate scales~\cite{Byrnes:2018txb}, see Fig.~\ref{fig:Pk}.

The running mass inflation model has a potential $V(\phi) = V_{0} + (1/2) m^2(\phi) \phi^2$, where the inflaton mass, $m$, depends on its value~\cite{Stewart:1996ey,Stewart:1997wg}. The resulting power spectrum can grow sufficiently for PBHs to form while satisfying constraints on cosmological scales~\cite{Leach:2000ea,Kohri:2007qn}. However this model is not complete; it relies on a Taylor expansion of the potential around a maximum and does not contain a mechanism for ending inflation
(see discussion in e.g. Sec. IV of Ref.~\cite{Motohashi:2017kbs}).

Inflation can alternatively be studied using the hierarchy of (Hubble) slow-roll parameters rather than the potential. It is possible to `design' a functional form for $\epsilon(N)$, where $N$
is the number of e-foldings of inflation, which satisfies all of the observational constraints and produces PBHs~\cite{Kohri:2007qn}. The corresponding potential has a `hill-top' form~\cite{Kohri:2007qn,Alabidi:2009bk}, with inflation occurring as the field evolves away from a local maximum, towards a minimum with $V(\phi) \neq 0$. 
However, as for running mass inflation, an auxiliary mechanism is required to terminate inflation, so this is not a complete single-field inflation model.

The reheating era at the end of inflation, where the inflaton oscillates around a minimum of its potential and decays, may offer a mechanism for generating PBHs~\cite{Green:2000he,Bassett:2000ha}. Reference~\cite{Martin:2019nuw} has shown that during oscillations about a parabolic minimum perturbations are enhanced sufficiently by a resonant instability for PBHs to be produced.

\subsubsection{Multi-field models}
\label{sec:multi}

In this subsection we discuss some multi-field scenarios which can generate large PBH-producing fluctuations: hybrid inflation, double inflation 
and a curvaton field. Quantum diffusion is also often important in multi-field models, e.g. Refs.~\cite{Randall:1995dj,GarciaBellido:1996qt,Yokoyama:1998pt,Clesse:2015wea}.

The most commonly studied two-field model in the context of PBH production is hybrid inflation (e.g.~Refs.~\cite{Randall:1995dj,GarciaBellido:1996qt,Lyth:2010zq,Clesse:2015wea}).
In hybrid inflation one of the fields, $\phi$, initially slow-rolls while the accelerated expansion is driven by the false-vacuum energy of a second scalar field $\psi$. At a critical value of $\phi$ there is a phase transition, with $\psi$ undergoing a waterfall transition to a global minimum and inflation ending. Around the phase transition quantum fluctuations are large, and a spike in the power spectrum on small scales is generated, leading to a large abundance of light PBHs~\cite{Randall:1995dj,GarciaBellido:1996qt}.
 For some parameter values, however, the waterfall transition can be `mild' so that there is a 2nd phase of inflation as the $\psi$ field evolves to the minimum of its potential~\cite{Clesse:2015wea}. In this case during the initial stage of the waterfall transition when both fields are important, isocurvature perturbations are generated, leading to a broad peak in the curvature perturbation power spectrum.  Perturbations on cosmological scales are generated during the initial phase of inflation and can be consistent with CMB observations. 

In double inflation~\cite{Silk:1986vc,Kawasaki:1997ju,Yokoyama:1998pt,Clesse:2015wea} there are two separate periods of inflation, with perturbations on cosmological scales being generated during the first period, and those on small scales during the second.~\footnote{In the initial realizations of double inflation~\cite{Silk:1986vc} different fields were responsible for the two periods of inflation. However the single field models with a local minimum or inflection point in the potential described in Sec.~\ref{sec:single} can also be viewed as double inflation models in the sense that the potential changes shape and there are two (or more) distinct dynamical phases of inflation~\cite{Kannike:2017bxn}.} Hybrid inflation models with a mild waterfall transition, as discussed above, fall into this class.

A curvaton is a field which is dynamically unimportant during inflation and
acquires isocurvature perturbations, with adiabatic perturbations being generated when it later decays~\cite{Lyth:2001nq}. If the inflaton is responsible for the perturbations on cosmological scales, while the curvaton generates small-scale perturbations, it is easier to produce large PBH-producing perturbations than in standard single field models (where the inflaton is responsible for perturbations on all scales)~\cite{Yokoyama:1995ex,Kawasaki:2012wr}. A specific example is the `axion-like curvaton'~\cite{Kawasaki:2012wr}.

\subsection{Other formation mechanisms}
\label{sec:otherform}

There are a variety of other early Universe processes which can produce large, PBH-forming overdensities. These include bubble collisions, cosmic string loop- or cusp-collapse, domain wall collapse and scalar condensate fragmentation.

First order phase transitions occur through the formation of bubbles. If these bubbles collide, PBHs with mass of order the horizon mass can form~\cite{Crawford:1982yz,Hawking:1982ga,Kodama:1982sf}. However a non-negligible abundance of PBHs
is only formed if the bubble formation rate is fine-tuned so that bubble collisions occur, but the phase transition doesn't complete instantaneously. Recently Ref.~\cite{Kusenko:2020pcg} has studied PBH formation from the collapse of bubbles nucleated during inflation.

Cosmic strings are topological defects which may form during phase transitions in the early Universe~\cite{Kibble:1976sj}.
A network of cosmic strings is formed which quickly reaches a stable scaling solution, in which loops with size smaller than the Hubble
radius are constantly being produced via long string interactions and loop self-intercommutations. The loops oscillate and if a loop contracts under its own tension to become smaller than its Schwarzschild radius a PBH can form~\cite{Hawking:1987bn,Polnarev:1988dh}.
The loop collapse probability is independent of time, and the mass of the PBH formed is proportional to the typical loop mass, which is proportional to the horizon mass. Consequently the PBHs formed from loop collapse have an extended mass function of the form ${\rm d} n_{\rm PBH}/ {\rm d} M_{\rm PBH} \propto M_{\rm PBH}^{-5/2}$~\cite{MacGibbon:1997pu}. The fraction of loops that collapse, $f$, is not well known. Numerical simulations~\cite{Caldwell:1991jj} have found $f = 10^{4.9 \pm 0.2} (G \mu)^{4.1 \pm 0.1}$ for large tensions  $ G \mu \sim 10^{-(2-3)}$.~\footnote{The stochastic gravitational wave background produced by loop oscillations now lead to a constraint $ G \mu < 1.5 \times 10^{-11}$~\cite{Blanco-Pillado:2017rnf}.}   
Critical phenomena also arise in this PBH formation mechanism, with the PBH mass scaling as a power law of the difference between the loop radius and the Schwarzschild radius~\cite{Helfer:2018qgv}. 
It has recently been argued that PBHs would form more abundantly from the collapse of cosmic string cusps~\cite{Jenkins:2020ctp}.

Large closed domain walls, produced during a second order phase transition, can collapse to form PBHs~\cite{Rubin:2000dq,Rubin:2001yw}.  The PBHs have masses that depend on the parameters of the field which undergoes the phase transition, typically with a significant spread, and can be clustered.

A scalar field with a sufficiently flat potential (such as the multiple flat directions found in supersymmetric generalizations of the Standard Model of particle physics) forms a coherent condensate at the end of inflation. This condensate typically fragments into lumps, 
such as oscillons or Q-balls. These lumps can come to dominate the Universe, and have large density fluctuations which can produce PBHs~\cite{Cotner:2016cvr,Cotner:2019ykd}.  These PBHs are smaller (compared with the horizon mass) than those formed via the collapse of density perturbations during radiation domination and can have larger spin~\cite{Cotner:2019ykd}.

Baryogenesis scenarios with spontaneous breaking of charge symmetry during inflation generate high density regions that can collapse to form PBHs after the QCD phase transition~\cite{Dolgov:1992pu}. In this case the PBHs formed have a lognormal mass function~\cite{Dolgov:1992pu}, centered at $\sim 10 \,M_{\odot}$ or higher~\cite{Dolghov:2020hjk}.

\section{Abundance constraints}
\label{sec:constraints}

PBH DM has a wide range of potentially observable effects. In this section we review the constraints on the present day abundance of PBHs, expressed as the fraction of DM in the form of PBHs today: $f_{\rm PBH}= \Omega_{\rm PBH}/\Omega_{\rm DM}$. 
We order the constraints roughly by increasing PBH mass:
evaporation (Sec.~\ref{sec:evap}),  interactions with stars (Sec.~\ref{sec:stars}), gravitational lensing (Sec.~\ref{sec:lensing}), gravitational waves from mergers of PBH binaries (Sec.~\ref{sec:gwmergers}), dynamical effects (Sec.~\ref{sec:dynamical}), the consequences of accretion (Sec.~\ref{sec:accretionconstraint}) and large scale structure (Sec.~\ref{sec:lss}).
We then review indirect constraints which apply if PBHs are formed via the collapse of large density perturbations during radiation domination (Sec.~\ref{sec:indirect}) and potential future constraints (Sec.~\ref{sec:future}).  For more detailed descriptions of the physics behind the constraints, including key equations, see Ref.~\cite{Sasaki:2018dmp}. 

\begin{figure}[t]
\begin{center}
\includegraphics[width=0.32\textwidth]{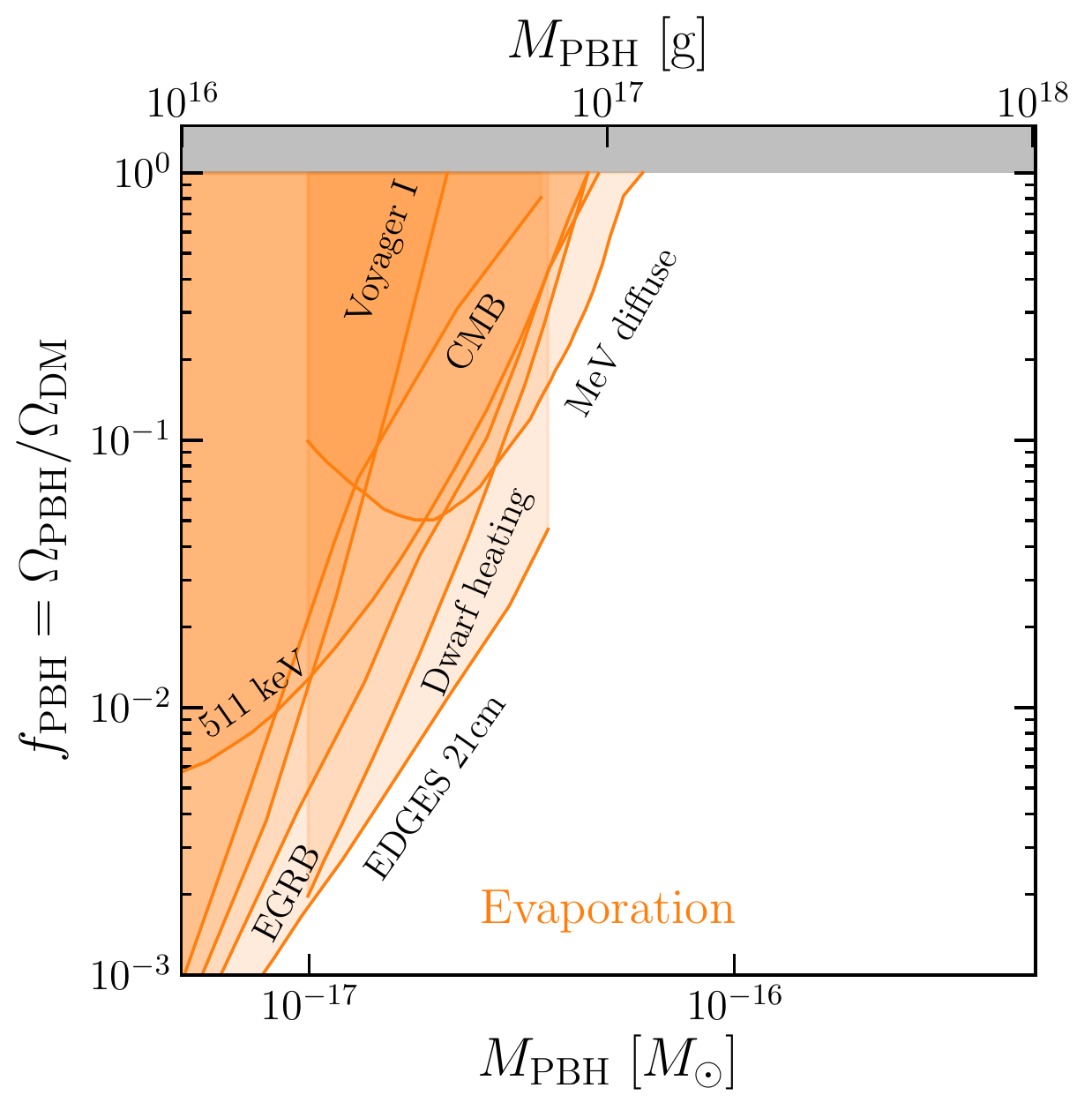}
\includegraphics[width=0.32\textwidth]{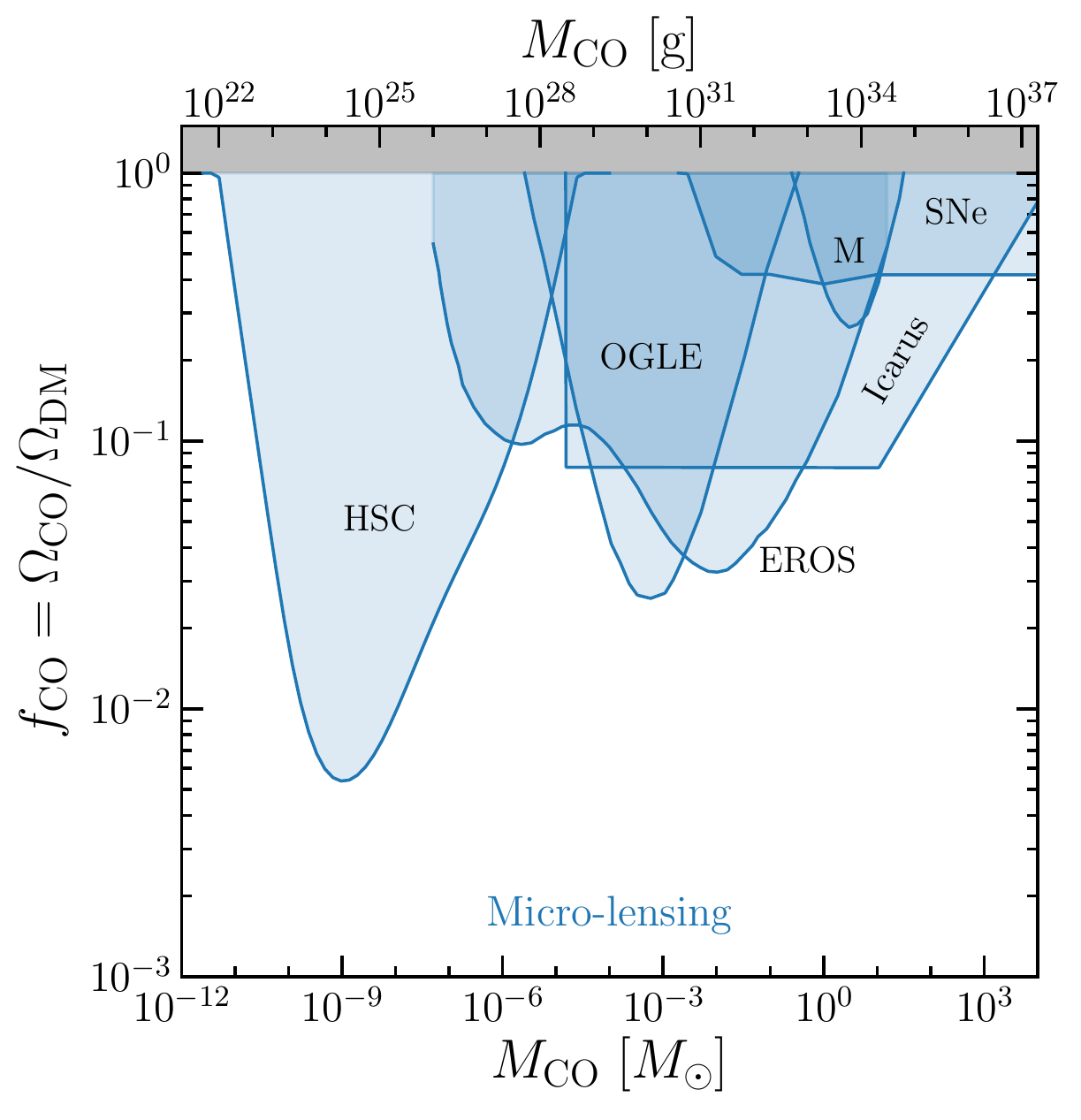}
\includegraphics[width=0.32\textwidth]{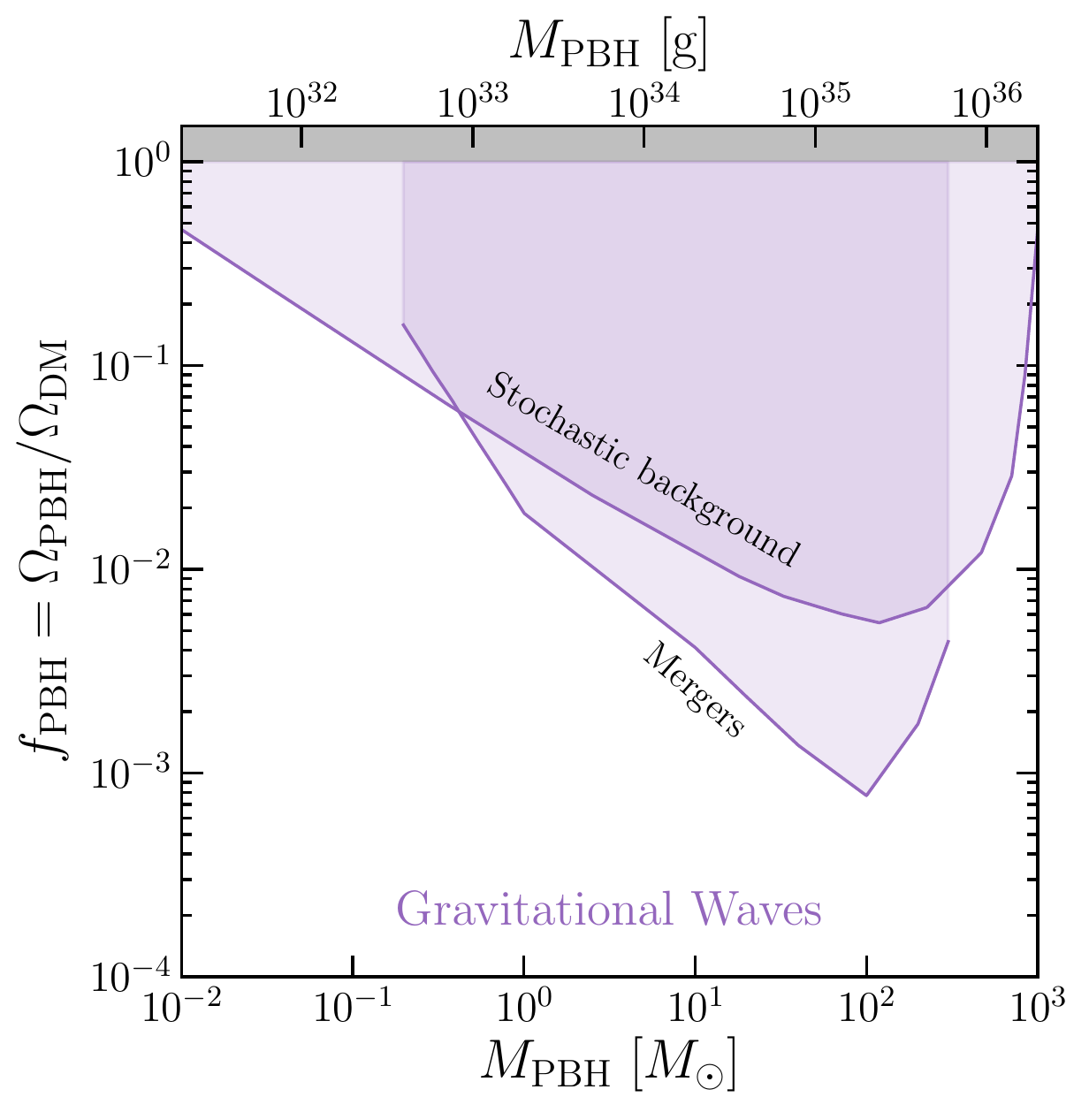}\\
\includegraphics[width=0.32\textwidth]{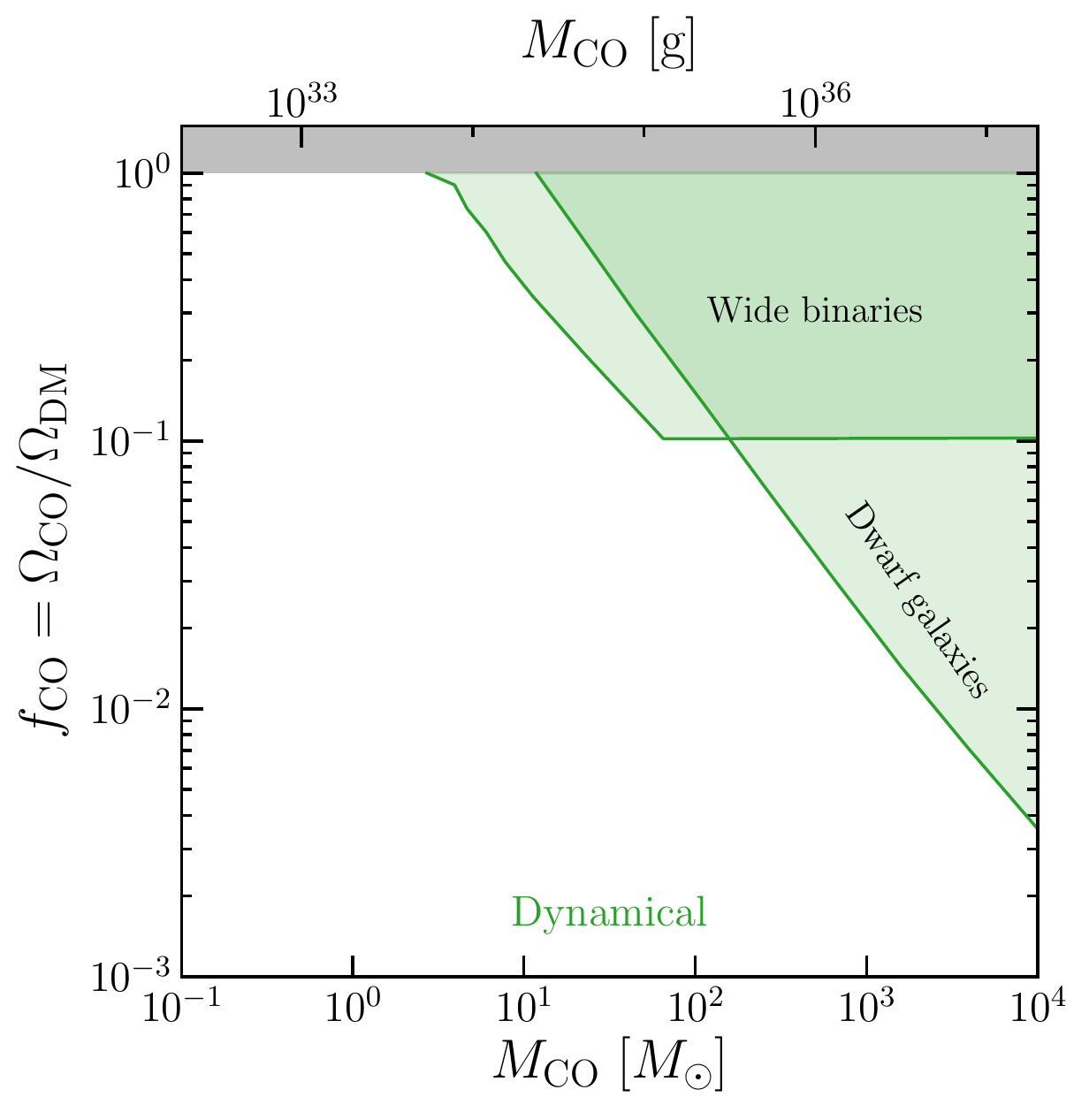}
\includegraphics[width=0.32\textwidth]{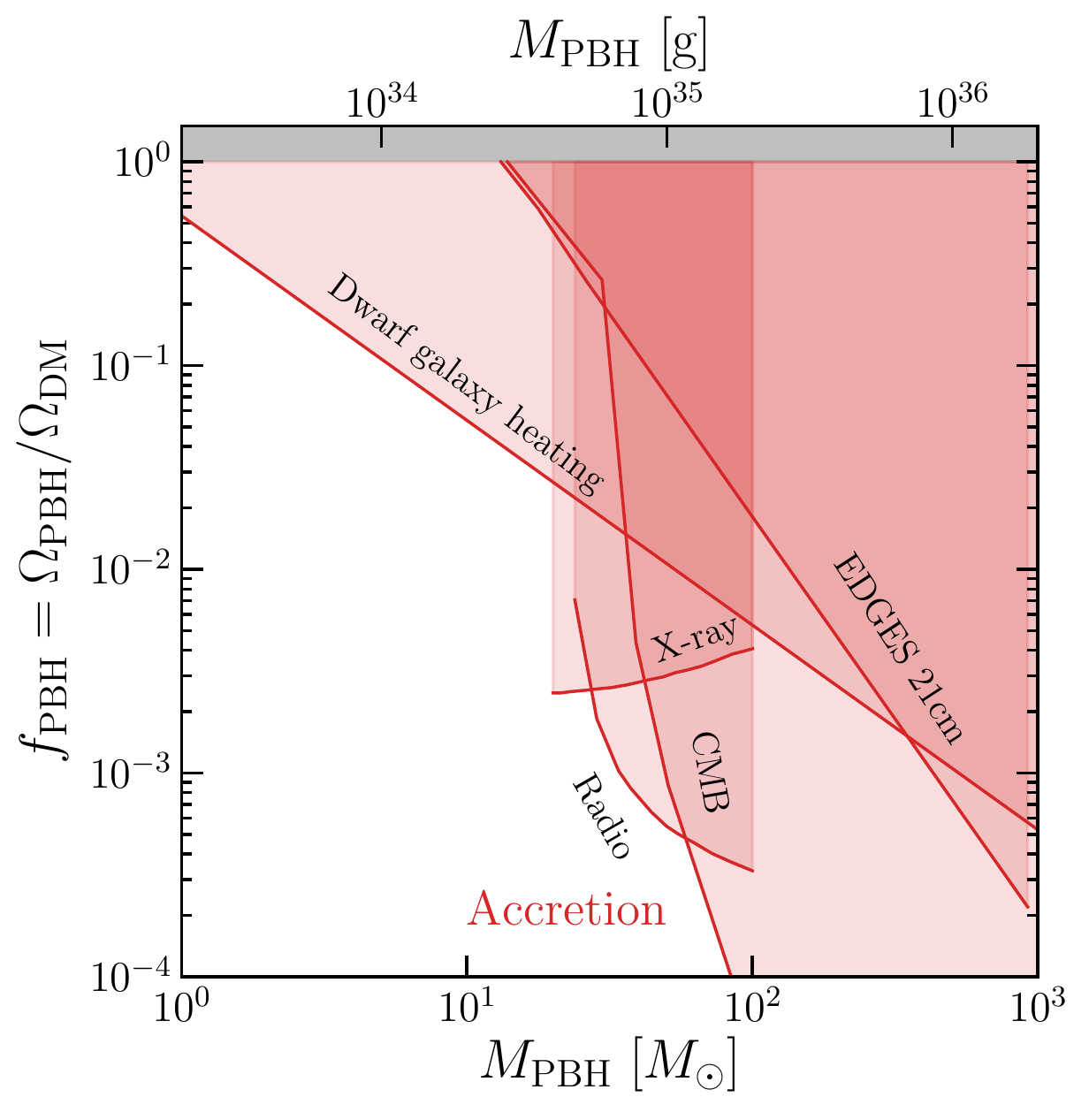}
\end{center}
\caption{Constraints on the fraction of DM in the form of PBHs $f_{\rm PBH}$, with mass $M_{\rm PBH}$, or in the form of compact objects, $f_{\rm CO}$, with mass $M_{\rm CO}$ for each of the different types of constraint. In each case the excluded regions are shaded. \textit{Top left:} Evaporation constraints on PBHs (Sec.~\ref{sec:evap}):  extragalactic gamma-ray background~\cite{Carr:2009jm}, CMB \cite{Poulin:2016anj,Clark:2016nst}, dwarf galaxy heating~\cite{Kim:2020ngi}, EDGES 21cm~\cite{Clark:2018ghm},  Voyager $e^{\pm}$~\cite{Boudaud:2018hqb}, $511 \, {\rm keV}$ gamma-ray line~\cite{DeRocco:2019fjq,Laha:2019ssq} and the MeV Galactic diffuse flux~\cite{Laha:2020ivk}. \textit{Top middle:} Gravitational lensing constraints on compact objects (Sec.~\ref{sec:lensing}): stellar microlensing (MACHO~\cite{Allsman:2000kg}, EROS~\cite{Tisserand:2006zx}, OGLE~\cite{Niikura:2019kqi}, HSC~\cite{Croon:2020ouk}), Icarus lensing event~\cite{Oguri:2017ock}, and supernovae magnification distribution~\cite{Zumalacarregui:2017qqd}. \textit{Top right:} 
Constraints on PBHs from gravitational waves (Sec.~\ref{sec:gwmergers}) produced by individual mergers~\cite{Kavanagh:2018ggo,Authors:2019qbw} and the stochastic background of mergers~\cite{Chen:2019irf}. Note that there are substantial uncertainties on GW constraints, arising from the possible disruption of PBH binaries. \textit{Bottom left:} Dynamical constraints on compact objects (Sec.~\ref{sec:dynamical}):  from dwarf galaxies~\cite{Brandt:2016aco} and wide binaries~\cite{mr}. 
\textit{Bottom right:} Accretion constraints on PBHs (Sec.~\ref{sec:accretionconstraint}):  CMB~\cite{Serpico:2020ehh}, EDGES 21cm~\cite{Hektor:2018qqw}, X-ray~\cite{Manshanden:2018tze}, radio~\cite{Manshanden:2018tze}, and dwarf galaxy heating~\cite{Lu:2020bmd}.
Digitised bounds and plotting codes are available online at \href{https://github.com/bradkav/PBHbounds} {\underline{PBHbounds}}.
}
\label{fig:constraints}
\end{figure}

\begin{figure}[t]
\begin{center}
\includegraphics[width=1.0\textwidth]{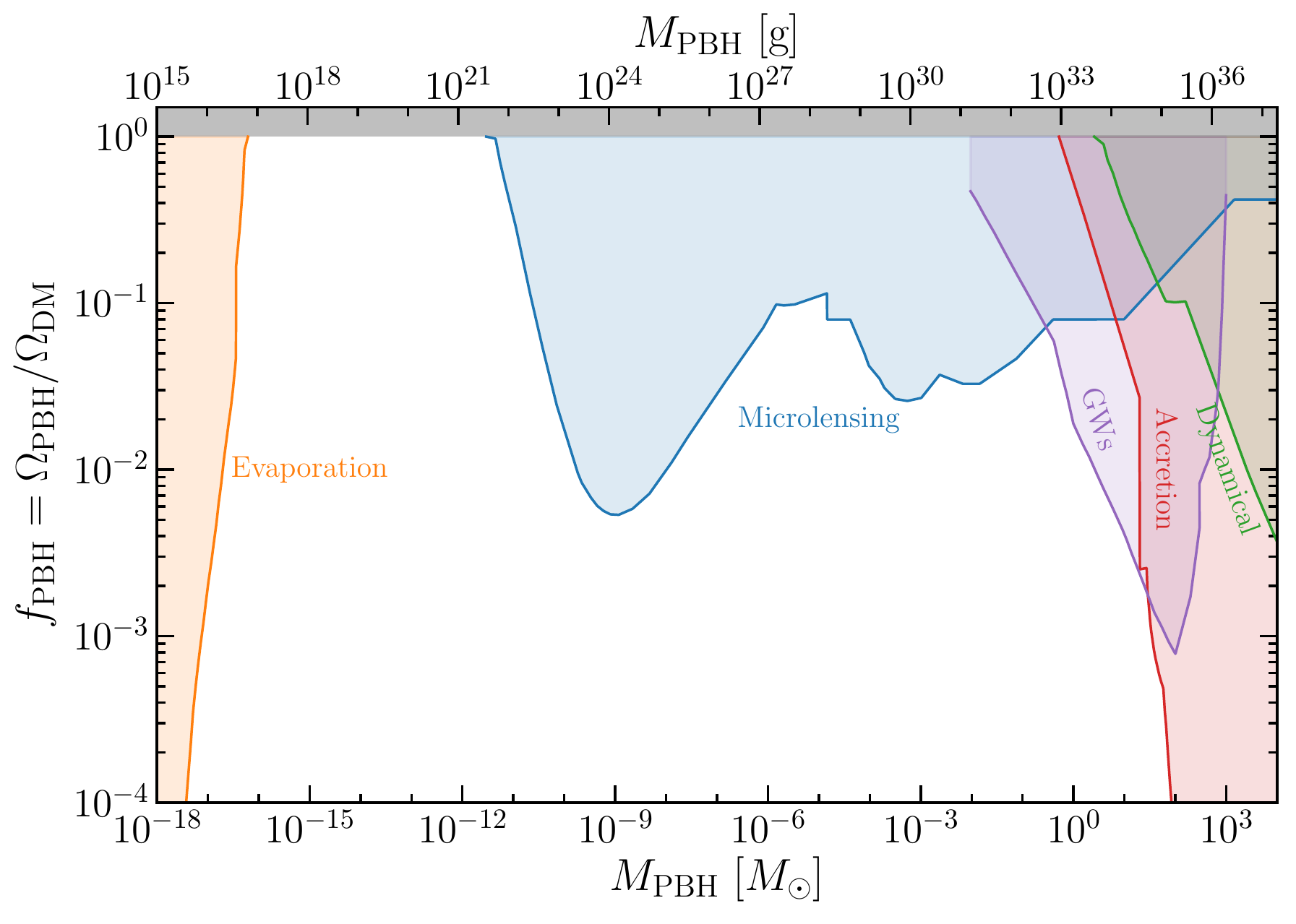}\\
\end{center}
\caption{\label{fig:allconstraints} All constraints on the fraction of DM in the form of PBHs, $f_{\rm PBHs}$, with mass $M_{\rm PBH}$, coming from PBH evaporation, microlensing, gravitational waves, PBH accretion and dynamical constraints. Each region shows the envelope of constraints from the corresponding panel in Fig.~\ref{fig:constraints}. Digitised bounds and plotting codes are available online at \href{https://github.com/bradkav/PBHbounds} {\underline{PBHbounds}}. }
\end{figure}

Figure~\ref{fig:constraints} shows all of the current limits discussed in the text, grouped by the type of constraint (evaporation, lensing, gravitational waves, dynamical effects, and accretion), while Fig.~\ref{fig:allconstraints} provides an overview, showing the envelope of each type of constraint. Where different constraints arise from different assumptions on e.g.~modeling and backgrounds, we aim to show the most conservative. Code for plotting the constraints is available online at \href{https://github.com/bradkav/PBHbounds}{github.com/bradkav/PBHbounds}.

We restrict our attention to PBHs with $M_{\rm PBH} \ll 10^{7} \, M_{\odot}$ which could, in principle, constitute the DM halos of small dwarf galaxies. There are various constraints on the abundance of more massive PBHs, for an overview, see Ref.~\cite{Carr:2020gox}. All limits quoted assume that the PBHs have a delta-function mass function and do not form clusters. We discuss the application of delta-function  constraints to extended mass functions in Sec.~\ref{sec:emf}. As discussed in Sec.~\ref{sec:gwmergers} understanding the late time clustering of PBHs is an outstanding challenge. In this section we use `PBH' to denote limits which apply specifically to PBHs and `CO' to denote limits which apply to any compact object.

\subsection{Evaporation}
\label{sec:evap}

PBHs with initial mass $M_{\rm PBH} < M_{\star} \approx 5 \times 10^{14} \, {\rm g}$ have completed their evaporation by the present day~\cite{Page:1976df,MacGibbon:2007yq}. The emission from slightly more massive PBHs ($M_{\star} < M_{\rm PBH} \lesssim 10^{17} \, {\rm g}$)  is sufficient that limits on their evaporation products can be used to constrain their abundance~\cite{Page:1976wx}. 

From the extragalactic gamma-ray background~\cite{Carr:2009jm,Carr:2016drx},
$f_{\rm PBH} \lesssim 2 \times 10^{-8} (M_{\rm PBH}/M_{\star})^{(3+ \epsilon)}$,
where  $\epsilon \sim 0.1-0.4$ parameterizes the energy dependence of the observed gamma-ray intensity: $I^{\rm obs} \propto E_{\gamma}^{-(1+\epsilon)}$\cite{Carr:2009jm}. This constraint can be tightened by a factor of ${\cal O}(10)$ by taking into account the contribution of known astrophysical sources, such as blazars~\cite{Ballesteros:2019exr}. There are also similar constraints from the 
damping of CMB anisotropies due to energy injection during recombination~\cite{Poulin:2016anj,Clark:2016nst}, from 
heating of neutral hydrogen, as probed by the EDGES measurements of 21cm absorption~\cite{Clark:2018ghm} and also from heating of the interstellar medium in dwarf galaxies~\cite{Kim:2020ngi}.

Constraints on the $e^{\pm}$ flux from Voyager 1 lead to a similar limit on the contribution of PBHs to the {\em local} dark matter density $f_{\rm PBH} < 0.001$ for $M_{\rm PBH} = 10^{16} \, {\rm g}$~\cite{Boudaud:2018hqb}, with the constraint on $f_{\rm PBH}$ varying with $M_{\rm PBH}$ in a similar way to the gamma-ray constraint. 
Again subtraction of backgrounds, in this case supernova remnants and pulsar wind nebulae, leads to constraints that are tighter by $\sim 2$ orders of magnitude~\cite{Boudaud:2018hqb}. 
Positrons produced by PBHs will also annihilate and contribute to the flux of the $511 \, {\rm keV}$  line~\cite{cg}.  The SPI/INTEGRAL limits on this line lead to limits on the PBH fraction which are similar to the gamma-ray limit for $10^{16} \, {\rm g} \lesssim M_{\rm PBH} \lesssim 10^{17} \, {\rm g}$~\cite{DeRocco:2019fjq,Laha:2019ssq}. There are somewhat tighter constraints (that exclude $f_{\rm PBH} = 1$ up to $M_{\rm PBH} \approx 2 \times 10^{17} \, {\rm g}$) from INTEGRAL measurements of the Galactic diffuse flux in the MeV range~\cite{Laha:2020ivk}. There are also constraints from Super-Kamiokande measurements of the diffuse neutrino background~\cite{Dasgupta:2019cae}.

\subsection{Interactions with stars}
\label{sec:stars}

Asteroid mass PBHs can potentially be constrained by the consequences of their capture by, and transit through, stars~\cite{Capela:2013yf,Pani:2014rca,Graham:2015apa,Montero-Camacho:2019jte}. See Ref.~\cite{Montero-Camacho:2019jte} for detailed recent calculations and discussion.

As a PBH passes through a star it loses energy by dynamical friction, and may be captured. A captured PBH will sink to the centre of the star and also accrete matter, potentially destroying the star. A large capture probability requires a large DM density and low velocity dispersions. Stellar survival constraints have been applied to globular clusters~\cite{Capela:2013yf}. However, as emphasised by Ref.~\cite{Pani:2014rca}, (most) globular clusters are not thought to have a high DM density.
 Moreover, Ref.~\cite{Montero-Camacho:2019jte} argues that the survival of stars does not in fact constrain the PBH abundance, but that the disruption of stars may lead to constraints, if the observational signatures are worked out (see Ref.~\cite{Genolini:2020ejw} for work in this direction). 
 
The transit of a PBH through a carbon/oxygen white dwarf will lead to localized heating by dynamical friction, which could ignite the carbon and potentially cause a runaway explosion~\cite{Graham:2015apa,Montero-Camacho:2019jte}.  Reference~\cite{Montero-Camacho:2019jte} again finds that the survival of white dwarfs does not constrain $f_{\rm PBH}$, but if white dwarf ignition by a PBH leads to a visible explosion there could be constraints.

\subsection{Gravitational lensing}
\label{sec:lensing}

\subsubsection{Stellar microlensing}
\label{sec:micro}

Stellar microlensing occurs when a compact object with mass in the range $5 \times 10^{-10} \, M_{\odot} \lesssim M_{\rm CO} \lesssim 10 \, M_{\odot}$ crosses the line of sight to a star, leading to a temporary, achromatic amplification of its flux~\cite{Paczynski:1985jf}. The duration of the microlensing event is proportional to $M_{\rm CO}^{1/2}$, therefore the range of masses constrained depends on the cadence of the microlensing survey. The EROS-2 survey of the Magellanic Clouds (MC) found $f_{\rm CO} \lesssim 0.1$ for masses in the range $10^{-6} \lesssim M_{\rm CO}/M_{\odot} \lesssim 1$. The constraint weakens above $M_\mathrm{CO} \approx 1\,M_\odot$, reaching $f_{\rm CO} \lesssim 1$ for $M_{\rm CO} \approx 30 M_{\odot}$~\cite{Tisserand:2006zx}. A MACHO collaboration search for long duration ($> 150$ days) events places a similar constraint on $f_{\rm CO}$, for $1 \lesssim M_{\rm CO}/M_{\odot} \lesssim 30$~\cite{Allsman:2000kg}. Uncertainties in the density and velocity distribution of the dark matter have a non-negligible effect on the MC microlensing constraints~\cite{Hawkins:2015uja,Green:2017qoa,Calcino:2018mwh}, and they would also be changed significantly if the compact objects are clustered~\cite{Calcino:2018mwh}.

Tighter constraints ($f_{\rm CO} \lesssim 10^{-2}$ for $M_{\rm CO} \sim 10^{-3} M_{\odot}$ weakening to $f_{\rm CO} \lesssim 0.1$ for $M_{\rm CO} \sim 10^{-5} \, M_{\odot}$ and $10^{-2} \, M_{\odot}$) have been obtained~\cite{Niikura:2019kqi} using the OGLE microlensing survey of the Galactic bulge~\cite{OGLE}. The OGLE data also contain 6 ultra-short ($0.1-0.3$ day) events which could be due to free-floating planets, or PBHs with $M_{\rm CO} \sim (10^{-4}-10^{-6}) M_{\odot}$ and $f_{\rm CO} \sim 0.01-0.1$~\cite{Niikura:2019kqi}.

A high cadence optical observation of M31 by Subaru HSC~\cite{Niikura:2017zjd} constrains $f_{\rm CO} \lesssim 10^{-2}$ for
$ 5 \times 10^{-10} \, M_{\odot} \lesssim M_{\rm CO} \lesssim  10^{-8} \, M_{\odot}$ weakening to $f_{\rm CO} \lesssim 1$ at 
$M_{\rm CO} \sim 5 \times 10^{-12} \, M_{\odot}$ and $ 5 \times 10^{-6} \, M_{\odot}$~\cite{Croon:2020ouk}. The constraints are weaker than initially found, due to finite source and wave optics effects. For $M_{\rm PBH} \lesssim 10^{-10} M_{\odot}$, the Schwarzschild radius of the PBH is comparable to, or less than, the wavelength of the light and wave optics effects reduce the amplification~\cite{Sugiyama:2019dgt}. Furthermore the stars in M31 that are bright enough for microlensing to be detected are typically larger than assumed in Ref.~\cite{Niikura:2017zjd},  further weakening the constraint~\cite{Montero-Camacho:2019jte,Smyth:2019whb}.
There are also significantly weaker constraints for $M_{\rm CO} \approx (10^{-7}-10^{-9}) \, M_{\odot}$ from a search for low amplification microlensing events in Kepler data~\cite{Griest:2013aaa}.

When a background star crosses a caustic in a galaxy cluster it is magnified by orders of magnitude~\cite{miraldaescude}. Microlensing by stars or other compact objects in the cluster can lead to short periods of further enhanced magnification. On the other hand if a significant fraction of the DM within the cluster is composed of compact objects then the overall magnification is reduced
(see e.g.~Ref.~\cite{Venumadhav:2017pps} and references therein). 
Icarus/MACS J1149LS1 is the first such microlensing event discovered (serendipitously) and is consistent with microlensing by an intracluster star of a source star at a redshift of 1.5~\cite{Kelly:2017fps,Diego:2017drh}. This leads to a constraint $f_{\rm CO} < 0.08$ for $ 10^{-5} < M_{\rm CO}/M_{\odot} < 10$  from the compact object population not reducing the magnification~\cite{Oguri:2017ock}. For more massive compact objects a constraint  $f_{\rm CO} < 0.08\,(M_{\rm CO}/10 \,M_{\odot})^{1/3}$ arises from assuming the microlensing event was caused by a dark compact object rather than a star 
~\cite{Oguri:2017ock}. Both of these constraints have an uncertainty of order a factor of $2$, from the uncertainty in the lens-source transverse velocity.

\subsubsection{Quasar microlensing}
Microlensing by compact objects in the lens galaxy leads to variations variation in the brightness of multiple-image quasars~\cite{Chang:1979zz}. Using optical data from 24 lensed quasars, Ref.~\cite{Mediavilla:2017bok} finds that $(20 \pm 5) \% $ of 
the mass of the lens galaxies is in compact objects (including stars) with mass in the range $0.05 < M_{\rm CO}/M_{\odot} < 0.45$. This is consistent with the expected stellar component, with only a small contribution allowed from dark compact objects, however no constraint on $f_{\rm CO}$ is stated.~\footnote{ According to Ref.~\cite{schechter} similar analysis of X-ray flux ratios, taking into account the stellar contribution, places a constraint $f_{\rm CO} \lesssim 0.1$.}

\subsubsection{Type 1a supernovae} The effects of gravitational lensing on the magnification distribution of type 1a supernovae (SNe1a) depend on whether or not the DM is smoothly distributed~\cite{Metcalf:2006ms}. If the DM is in compact objects with $M_{\rm CO}  \gtrsim 10^{-2} M_{\odot}$, then most SNe1a would be dimmer than if the DM were smoothly distributed, while a few will be significantly magnified~\cite{Zumalacarregui:2017qqd}. Using the JLA and Union 2.1 SNe samples, Ref.~\cite{Zumalacarregui:2017qqd} find a constraint $f_{\rm CO} \lesssim 0.4$, for all $M_{\rm CO}  \gtrsim 10^{-2} M_{\odot}$. They argue that this result is robust to the finite size of SNe and peculiar SNe. The former is contrary to Ref.~\cite{Garcia-Bellido:2017imq} who argue that the constraint does not apply for $M \lesssim 3 M_{\odot}$. 

\subsubsection{Strong lensing of Fast Radio Bursts}

Strong gravitational lensing of Fast Radio Bursts (FRBs) by COs with $M_{\rm CO} \gtrsim (10-100) M_{\odot}$ would lead to two images, separated by a measurable (ms) time delay~\cite{Munoz:2016tmg}. No such signal has been seen in the $\sim 100$ FRBs observed to date, which leads to a constraint $f_{\rm CO} \lesssim 0.7$ for $M_{\rm CO} \gtrsim 10^{3} M_{\odot}$, weakening to $f_{\rm CO} \lesssim 1$ for $M_{\rm CO} \sim 10^{2} M_{\odot}$~\cite{Liao:2020wae}.

\subsubsection{Femtolensing}
\label{femto}
Reference~\cite{gould} proposed that asteroid mass compact objects 
could be probed by femtolensing of gamma-ray bursts (GRBs), specifically via interference fringes in the frequency spectrum due to the different phases of the 2 (unresolved) images during propagation. 
Using data from the Fermi Gamma-ray Burst Monitor, Ref.~\cite{Barnacka:2012bm} placed constraints on compact objects in the mass range $ 5 \times 10^{17} \, {\rm g} \lesssim M_{\rm CO} <  10^{20} \, {\rm g}$. However Ref.~\cite{Katz:2018zrn} demonstrated that most GRBs are too large to be modelled as point-sources, and furthermore that wave optics effects~\cite{Ulmer:1994ij} need to be taken into account. Consequently there are in fact no current constraints on $f_{\rm CO}$ from femtolensing.

\subsection{Gravitational waves from mergers}
\label{sec:gwmergers}

In the late 1990s Nakamura et al.~\cite{Nakamura:1997sm,Ioka:1998nz} studied the formation of Solar mass PBH DM binaries in the early Universe, when pairs of PBHs may be close enough to decouple from the Hubble expansion before matter-radiation equality.  Three-body interactions would impart a small angular momentum on the PBHs, leading to the formation of highly eccentric binaries.
If these binaries survive unaffected to the present day then the gravitational waves (GWs) resulting from their coalescence could be detected by LIGO. \footnote{An additional but sub-dominant contribution to the PBH merger rate comes from binaries formed by dynamical capture in the late Universe~\cite{Bird:2016dcv}.} In fact the merger rate would be several orders of magnitude larger than measured by LIGO-Virgo~\cite{Abbott:2016blz}, which places a tight constraint, $f_{\rm PBH} < {\cal O} (10^{-3})$ for $10 \lesssim M_{\rm PBH}/M_{\odot} \lesssim 300$~\cite{Hayasaki:2009ug,Sasaki:2016jop,Ali-Haimoud:2017rtz,Kavanagh:2018ggo}. A dedicated LIGO-Virgo search for sub-Solar mass mergers constrains  $f_{\rm PBH} < {\cal O} (10^{-1})$ down to $M_\mathrm{PBH} \sim 0.2\,M_\odot$ \cite{Authors:2019qbw}. There are also similar constraints from the stochastic gravitational wave background~\cite{Wang:2016ana,Raidal:2017mfl,Chen:2019irf,LIGOScientific:2019vic} of such mergers, as well as searches for PBH binaries with large mass ratios~\cite{Nitz:2020bdb}.

If PBHs don't constitute all of the DM, then during matter domination stellar mass (and more massive) PBHs accrete halos of particle dark matter with a steep density profile~\cite{Mack:2006gz,Adamek:2019gns}.~\footnote{Consequently stellar mass PBHs and WIMP DM can't coexist, as gamma-rays from WIMP annihilation in the WIMP halos around PBHs would have already been observed~\cite{Lacki:2010zf,Adamek:2019gns,Bertone:2019vsk}.} 
These DM mini-halos affect the dynamical evolution of the PBH-binaries~\cite{Hayasaki:2009ug,Ali-Haimoud:2017rtz}, however this has a relatively small effect on the merger rate and resulting constraints~\cite{Kavanagh:2018ggo}.

A major outstanding problem in the calculation of GW constraints is the evolution, and survival, of PBH binaries between formation and merger. If PBHs make up a significant fraction of the DM then PBH clusters form not long after matter-radiation equality~\cite{Chisholm:2005vm,Chisholm:2011kn,Raidal:2018bbj,Inman:2019wvr}. While distant three-body interactions are expected to have little impact on isolated PBH binaries~\cite{Ali-Haimoud:2017rtz,Young:2020scc}, three-body interactions in PBH clusters could significantly affect the properties of binaries and hence the predicted merger rates~\cite{Vaskonen:2019jpv,Jedamzik:2020ypm,Jedamzik:2020omx,Trashorras:2020mwn}. Merger rates may also be increased (leading to stronger constraints) in scenarios where PBHs are formed with large initial clustering (as discussed in Sec.~\ref{sec:spincluster})~\cite{Ballesteros:2018swv,Bringmann:2018mxj}, though Ref.~\cite{Atal:2020igj} argue that clustering instead should weaken constraints. Given these outstanding problems of PBH clustering and binary survival, in Fig.~\ref{fig:constraints} we show constraints~\cite{Kavanagh:2018ggo,Authors:2019qbw,Chen:2019irf} that assume no clustering and no disruption of the binaries.~\footnote{Note that the constraints in Ref.~\cite{Kavanagh:2018ggo} are derived using limits from LIGO's first observing run (O1). Stronger constraints can be derived from more recent data (O2), assuming that none of the observed binary BH mergers have a primordial origin~\cite{Vaskonen:2019jpv,Chen:2019irf}.}


\subsection{Dynamical}
\label{sec:dynamical}
\subsubsection{Dwarf galaxies}

Two-body interactions lead to the kinetic energies of different mass populations within a system becoming more equal. In a system made of stars and more massive compact objects, the stars gain energy and their distribution will expand. Ultra-faint dwarf galaxies (UFDGs) are particularly sensitive to this effect, due to their high mass to luminosity ratios. Reference~\cite{Brandt:2016aco} found that the observed sizes of UFDGs place a constraint $f_{\rm CO} \lesssim  0.002-0.004$ for $M_{\rm CO} \sim 10^{4} M_{\odot}$, weakening with decreasing mass to  $f_{\rm CO} \lesssim 1 $ for $M_{\rm CO} \sim 10 M_{\odot}$. The uncertainty comes from uncertainties in the  velocity dispersion of the compact objects and the assumed initial radius of the stellar distribution. Tighter constraints may be obtained from the survival of the star cluster near the centre of the dwarf galaxy Eridanus II, depending on its age~\cite{Brandt:2016aco}. 

Reference~\cite{Koushiappas:2017chw} used the projected stellar surface density profile of Segue 1, and in particular the absence of a ring feature, to show that $f_{\rm CO} < 0.06 \, (0.20)$ for $M_{\rm CO} = 30 \, (10) M_{\odot}$ at 99.9\% confidence. 
Subsequent  Fokker-Planck simulations of dwarfs composed of stars and compact object dark matter have not found such a feature however~\cite{Zhu:2017plg}, and have also found a slightly slower growth of the stellar component than in Ref.~\cite{Brandt:2016aco}. Consequently their constraints~\cite{Zhu:2017plg} are slightly weaker than Refs.~\cite{Brandt:2016aco,Koushiappas:2017chw}: $f_{\rm CO} < 1$ for $M_{\rm CO} >14 \, M_{\odot}$.  Reference~\cite{Stegmann:2019wyz} has subsequently shown that collectively UFDGs exclude $f_{\rm CO} = 1$ for the mass range $(1-100) M_{\odot}$ for both delta-function and lognormal mass functions.  While some low mass UFDGs have stellar populations that are individually consistent with $f_{\rm CO} \sim 1$ for $M_{\rm CO} \sim 1 M_{\odot}$ (see also Ref.~\cite{Zhu:2017plg}), most would be `puffed up' too much.

\subsubsection{Wide binaries}

The energy of wide binary stars is increased by multiple encounters with compact objects, potentially leading to disruption~\cite{bht}.
The separation distribution of wide binaries can therefore be used to constrain the abundance of compact objects~\cite{Yoo:2003fr}. Radial velocity measurements are required to confirm that wide binaries are genuine, as otherwise erroneously tight constraints can be obtained from spurious binaries~\cite{Quinn:2009zg}.

Using the 25 wide binaries in their catalogue~\cite{Allen}  that spend the least time in the Galactic disk (and are hence least affected by encounters with the stars therein) Ref.~\cite{mr} finds $f_{\rm CO} \lesssim 0.1$ for $M_{\rm CO} \gtrsim 70 M_{\odot}$ with the limit weakening with decreasing mass to $f_{\rm CO} \lesssim 1 $ for $M_{\rm CO} = 3 M_{\odot}$. Tighter constraints may be possible using data from \textit{Gaia}, however radial velocity follow-up will be needed to confirm that candidate binaries are genuine (c.f. Ref.~\cite{pwos}).

\subsection{Accretion}
\label{sec:accretionconstraint}

\subsubsection{$z>0$}
Radiation emitted due to gas accretion onto PBHs can modify the recombination history of the Universe, and hence affect the anisotropies and spectrum of the CMB~\cite{1981MNRAS.194..639C,Ricotti:2007au}. There are significant theoretical uncertainties in the accretion rate and also the ionizing effects of the radiation~\cite{Ali-Haimoud:2016mbv,Bosch-Ramon:2020pcz}. 

Reference~\cite{Poulin:2017bwe} argues that, contrary to the spherical accretion assumed in previous work, an accretion disk should form, in which case the resulting constraints are significantly tightened. Formation of (non-PBH) dark matter halos around PBHs tightens the constraints for $M_{\rm PBH} \gtrsim 10 M_{\odot}$~\cite{Serpico:2020ehh}. For spherical (disk) accretion $f_{\rm PBH} \lesssim 1$ for $M_{\rm PBH} \sim 10 \, (1) \, M_{\odot}$, tightening with increasing PBH mass to $f_{\rm PBH} < 3 \times 10^{-9}$ at $M_{\rm PBH} \sim 10^{4} M_{\odot}$~\cite{Serpico:2020ehh}. There are also model-dependent constraints on PBHs with $M_{\rm PBH}  \gtrsim 10 \, M_{\odot}$  from their effects on the 21-cm spectrum as measured by EDGES~\cite{Hektor:2018qqw}.

\subsubsection{Present day}

Accretion of interstellar gas onto $M_{\rm PBH}> M_{\odot}$ PBHs in the Milky Way would lead to observable X-ray and radio emission~\cite{Gaggero:2016dpq}. Comparing predictions from numerical simulations of gas accretion onto isolated moving compact objects with known X-ray and radio sources in the Chandra and VLA Galactic centre surveys leads to a constraint $f_{\rm PBH} \lesssim 10^{-3}$ for $M_{\rm PBH} \sim (30-100) \, M_{\odot}$~\cite{Manshanden:2018tze}. Reference~\cite{Inoue:2017csr} uses the observed number density of compact X-ray objects to place a similar constraint on $f_{\rm PBH}$, valid up to $M_{\rm PBH} \sim 10^{7} \, M_{\odot}$. Reference~\cite{Lu:2020bmd} places a constraint $f_{\rm PBH} \lesssim 10^{-4}$ for $M_{\rm PBH} \sim 10^{3} M_{\odot}$, weakening to 
$f_{\rm PBH} \lesssim 1$ for $M_{\rm PBH} \sim M_{\odot}$ and $10^{7} M_{\odot}$, from gas heating in dwarf galaxies.

\subsection{Large scale structure}
\label{sec:lss}

If massive PBHs make up a significant fraction of the DM, then Poisson fluctuations in their number density enhance the matter power spectrum at small scales~\cite{1975A&A....38....5M,Afshordi:2003zb}, which can be probed by observations of the Lyman-$\alpha$ forest~\cite{Afshordi:2003zb}. Using the latest data from MIKE/HIRES and high-resolution hydrodynamical simulations Ref.~\cite{Murgia:2019duy} finds a conservative limit $f_{\rm co} \lesssim (100 M_{\odot}/ M_{\rm PBH})$.

\subsection{Indirect constraints}
\label{sec:indirect}

In this subsection we look at constraints on the amplitude of large primordial perturbations, which lead to indirect constraints on the abundance of PBHs formed via the collapse of large density perturbations during radiation domination (Sec.~\ref{sec:raddom}). These constraints do not apply to PBHs formed via other mechanisms (see Sec.~\ref{sec:otherform}). As discussed in Sec.~\ref{sec:raddom}, there are large uncertainties in the calculation of the abundance of PBHs formed from a given primordial power spectrum.

First order scalar perturbations generate tensor perturbations at second order~\cite{Ananda:2006af,Baumann:2007zm}. If the density perturbations are sufficiently large then the amplitude of the resulting `scalar induced gravitational waves' (SIGWs) is larger than that of the GWs generated by the primordial tensor perturbations. Constraints on the energy density of stochastic GWs, from e.g.~Pulsar Timing Arrays, therefore limit the abundance of PBHs formed via the collapse of large density perturbations~\cite{Saito:2008jc}. These constraints depend on the shape of the primordial power spectrum, and also the assumed probability distribution of the density perturbations, and are therefore (inflation) model dependent~\cite{Garcia-Bellido:2016dkw,Inomata:2016rbd,Orlofsky:2016vbd}. Models which produce a broad peak in the primordial power spectrum are most tightly constrained~\cite{Inomata:2016rbd,Orlofsky:2016vbd}. 
For PBHs forming from large density perturbations during radiation domination, Refs.~\cite{Byrnes:2018txb,Inomata:2018epa} find $f_{\rm PBH} < 1$ for $10^{-2} \lesssim M_{\rm PBH}/M_{\odot} \lesssim 1$. Reference~\cite{Chen:2019xse} finds, using data from NANOGrav, $f_{\rm PBH} < 10^{-23}$ for $M_{\rm PBH} = 0.1 M_{\odot}$  and $f_{\rm PBH}  < 10^{-6}$ for $0.002 < M_{\rm PBH}/M_{\odot} < 0.4$. However this calculation makes approximations which have a huge effect on the constraint on $f_{\rm PBH}$ (including setting the PBH formation threshold equal to unity, and $\sigma^2 = A$). There are also tight constraints on the abundance of light, $M_{\rm PBH} \sim 10^{13-15} \, {\rm g}$, PBHs from limits on SIGWs from LIGO~\cite{Kapadia:2020pir}. Such light PBHs are expected to have evaporated by the present day, however if Hawking evaporation is not realised in nature they would be stable and otherwise viable as DM.

The amplitude of the primordial density perturbations can also be constrained by the CMB spectral distortions caused by the dissipation of large perturbations~\cite{Carr:1993aq}. Limits on CMB spectral distortions from COBE/FIRAS exclude $f_{\rm PBH} = 1$ for PBHs with 
$ M_{\rm PBH}/M_{\odot} \gtrsim 10^{3}$
formed from large density perturbations with a Gaussian distribution~\cite{Kohri:2014lza}.~\footnote{Here we have used Eq.~(\ref{mhr}), taken from Ref.~\cite{Wang:2019kaf}, to translate from $k$ to $M_{\rm PBH}\approx M_{\rm H}$.}
Masses down to $M \sim 10^{3} M_{\odot}$ can similarly be excluded via the effects of dissipation 
on Big Bang Nucleosynthesis~\cite{Inomata:2016uip}.

\subsection{Future constraints}
\label{sec:future}

In this subsection we discuss potential future constraints on PBH DM. We start with direct constraints in (roughly) increasing order of PBH mass probed, followed by indirect constraints.

Planned space observatories, such as e-ASTROGAM and ASTRO-H, will allow a lower and more precise measurement of the flux of the isotropic gamma-ray and X-ray backgrounds. This will allow improved constraints to be placed on PBHs in the mass range $M_{\rm PBH} \sim 10^{16-18} \, {\rm g}$ via the products of their evaporation~\cite{Ballesteros:2019exr}.

A small subset of GRBs with fast variability have small sizes and are therefore suitable targets for femtolensing~\cite{Katz:2018zrn}. A sample of 100 such GRBs with well-measured redshifts could be used to probe PBHs with $10^{17} \, {\rm g} \lesssim M_{\rm CO} \lesssim 10^{19} \, {\rm g}$. Proposals to measure the lensing parallax of GRBs -- the relative brightness measured by multiple telescopes at large spatial separations -- suggest that  this approach could be sensitive to the entire unconstrained range $M_\mathrm{PBH} \sim 10^{17-23}\,\mathrm{g}$~\cite{Nemiroff:1995ak,Jung:2019fcs}. Very high cadence observations of white dwarfs in the LMC, by e.g.~LSST, could reduce the minimum mass probed by microlensing observations by a factor of a few~\cite{Sugiyama:2019dgt}. However, as discussed in detail in Ref.~\cite{Montero-Camacho:2019jte}, diffraction and finite source size effects make it difficult to extend the range of masses probed by optical microlensing below $M_{\rm CO} \sim 10^{22} \, {\rm g}$. Microlensing of X-ray pulsars, which have small source sizes, can avoid these restrictions and long observations by X-ray telescopes with large effective areas (e.g.\ AstroSat, LOFT) of SMC X-1 and other X-ray binaries could probe the range $10^{18} \, {\rm g} \lesssim M_{\rm CO} \lesssim 10^{22} \, {\rm g}$~\cite{Bai:2018bej}.

Per-cent level constraints on $f_{\rm CO}$ for $10^{-4} \lesssim M_{\rm CO}/M_{\odot} \lesssim 0.1$ could be achieved from future FRB detections, via the phase difference they produce between unresolved images (as in femtolensing)~\cite{Katz:2019qug}. Pulsar timing arrays can detect the gravitational redshift and acceleration induced by passing compact objects~\cite{Schutz:2016khr,Dror:2019twh}. Via various types of searches, SKA will be able to probe $f_{\rm PBH} \approx 1$ over the entire mass range $(10^{-12} -100)\, M_{\odot}$, with potentially sub-percent level constraints in some mass regions~\cite{Dror:2019twh}.  Planned sub-Hertz gravitational wave observatories such as LISA and DECIGO will be sensitive to extreme mass ratio binaries, composed of astrophysical supermassive black holes and compact objects in the range $10^{-6} \lesssim M_{\rm CO}/M_{\odot} \lesssim 1$, down to the level of around $f_\mathrm{CO} \sim 10^{-3}$~\cite{Guo:2017njn,Wang:2019kzb}.

Detailed observations of caustic-crossing events (combined with improved theoretical modelling) could place very tight constraints on compact objects with planetary, stellar and larger masses~\cite{Diego:2017drh,Venumadhav:2017pps}.
Constraints on frequency-dependent gravitational lensing dispersions by future gravitational wave detectors could probe PBHs with $M \gtrsim 0.1 M_{\odot}$ via their effects on the matter power spectrum~\cite{Oguri:2020ldf}. Measurements of the 21cm power spectrum by HERA and SKA will potentially improve cosmological accretion constraints on PBHs with $M_{\rm PBH}> M_{\odot}$ by an order of magnitude~\cite{Mena:2019nhm}. 
Accurate astrometric surveys can probe compact objects via the time-dependent weak lensing of stars (``astrometric microlensing")~\cite{Dominik:1998tn}.
By the end of its lifetime {\it Gaia} could place sub per-cent level constraints on $f_{\rm CO}$ for $M_{\rm CO}> 10 \, M_{\odot}$ via the large anomalous angular velocities and accelerations produced by close encounters~\cite{VanTilburg:2018ykj}. Similar constraints could be placed on stellar and planetary mass COs by {\it Gaia} via non-repeating proper motion anomalies (dubbed `blips'), with a future survey such as {\it Theia} capable of even tighter constraints~\cite{VanTilburg:2018ykj}. 

Upcoming experiments, suchs as CHIME and SKA, should lead to constraints in the range $f_{\rm CO}< (0.01-0.1)$ for  $M_{\rm CO} \gtrsim (10-100) M_{\odot}$ from strong gravitational lensing of FRBs by COs~\cite{Munoz:2016tmg,Liao:2020wae}. 
Similar constraints could be obtained down to $M_{\rm CO} \approx 2 M_{\odot}$ from lensing of the burst microstructure~\cite{Laha:2018zav}. Strong lensing of GRBs by compact objects with $10 \lesssim M_{\rm CO}/M_{\odot} \lesssim 1000 M_{\odot}$ leads to superimposed images which could be detected by a future GRB observatory, leading to per-cent level constraints~\cite{Ji:2018rvg}.  
Gravitational lensing of gravitational waves by compact objects with $M_{\rm CO}/M_{\odot} \gtrsim 5 M_{\odot}$ would produce fringes which could lead to sub per-cent level constraints from current gravitational wave detectors~\cite{Diego:2019rzc} and 3rd generation experiments like the Einstein Telescope~\cite{Liao:2020hnx}.

Future space-based laser interferometers will be able to indirectly constrain PBHs produced by the collapse of large density perturbations in the mass range $10^{20-26} \, {\rm g}$, via induced gravitational waves~\cite{Saito:2008jc,Cai:2018dig,Bartolo:2018evs,Inomata:2018epa}. A PIXIE-like CMB spectral distortion experiment could similarly constrain $M_{\rm PBH} \gtrsim \, M_{\odot}$~\cite{Chluba:2019nxa}.

\subsection{Application of constraints to extended mass functions}
\label{sec:emf}
The constraints described above are typically calculated assuming a delta-function (or monochromatic) mass function. However, as discussed in Sec.~\ref{sec:dp}, in many cases PBHs are expected to be produced with an extended mass function.

Carr et al.~\cite{Carr:2017jsz} devised a method for applying constraints calculated assuming a delta-function mass function to specific extended mass functions, without explicitly recalculating the constraint from scratch. The dependence of a given astrophysical observable, $A[\psi(M)]$, on the mass function, $\psi(M)$, defined so that
$f_{\rm PBH} = \int \psi(M) \, {\rm d} M$
can be expanded as
\begin{equation}
A[\psi(M)] = A_{0} + \int \psi(M) K_{1}(M) \, {\rm d} M + \int \psi(M) K_{2}(M_{1}, M_{2}) \, {\rm d} M_{1} \,  {\rm d} M_{2}  + ...  \,,
\label{apsi}
\end{equation}
where $A_{0}$ is the background contribution and the functions $K_{j}(M)$ encode how the underlying physics of the observable depend on the PBH mass function. In many cases observations place a bound on a single observable: $A[\psi(M)] < A_{\rm exp}$. For instance in the case of stellar microlensing the observable is the number of events. Also, for many constraints, PBHs with different masses contribute independently to the constraint so that $K_{j}(M) =0$ for $j \geq 2$. In this case the constraint for a delta-function mass function as a function of mass,  $f_{\rm max}(M)$, can be translated into a constraint on a specified extended mass function using:
\begin{equation}
\int \frac{ \psi(M)}{f_{\rm max} (M)} \, {\rm d} M \leq 1 \,.
\end{equation}
This procedure has to be implemented separately for each observable, and different constraints combined in quadrature. As emphasised in Ref.~\cite{Carr:2017jsz} there are some caveats in the application of this method. The PBH MF can evolve, due to mergers and accretion, so that the the initial MF is not the same as the MF at the time the constraint applies. 
For some constraints, e.g.~GWs from mergers, the higher order terms, $K_{j}(M)$ for $j \geq 2$, are not zero and a detailed calculation is required for each mass function~\cite{Raidal:2017mfl,Chen:2018czv}. In some cases, such as the effects of PBH accretion on the CMB, the observable is not a single quantity, and if this is not taken into account artificially tight constraints will be obtained. 

This method has also been expounded on by Bellomo et al.~\cite{Bellomo:2017zsr}. They showed that for any specific mass function, for each observable the effects are equivalent to a delta-function MF with a particular `equivalent mass'. They also emphasised that when considering MFs with extended tails, for instance a lognormal distribution, care should be taken not to apply constraints beyond the limits of their validity.

When applied to extended mass functions the constraints are effectively `smeared out'~\cite{Carr:2017jsz}. Consequently, when multiple constraints are considered, extended mass functions are more tightly constrained than the delta-function mass function, and small mass windows between constraints are closed~\cite{Green:2016xgy,Carr:2017jsz} 

\section{Summary}
\label{sec:summary}

The LIGO-Virgo discovery of gravitational waves from multi-Solar mass black holes has led to a resurgence of interest in primordial black holes (PBHs) as a dark matter candidate. Consequently there have been significant improvements in the theoretical calculations of PBH formation and also the observational constraints on their abundance. In this final section we summarise the current status and highlight key open questions. 

The most popular PBH formation mechanism is the collapse of large density perturbations, generated by a period of inflation in the early Universe, during radiation domination. There have been significant recent improvements in our understanding of the threshold for collapse, $\delta_{\rm c}$, and its dependence on the shape of the perturbations (see Sec.~\ref{sec:deltac}). To produce an interesting number of PBHs the primordial power spectrum must grow by $\sim 7$ orders of magnitude from its measured value on cosmological scales. This can be achieved in single-field models with an inflection point or a shallow local minimum in the potential (Sec.~\ref{sec:single}), or in some multi-field models (Sec.~\ref{sec:multi}), and there has been significant recent work revisiting these models and refining calculations. The standard calculation of the abundance and mass function of PBHs (Sec.~\ref{sec:betamf}) assumes that the primordial density perturbations have a Gaussian probability distribution. However this assumption is not valid for large, PBH-producing perturbations. An accurate calculation of the non-Gaussian probability distribution, which for many models needs to take into account quantum diffusion, is an open issue. Detailed calculations of the mass function and initial clustering of PBHs produced by broad power spectra are also an outstanding problem.

There are a wide range of different observational constraints. The lightest stable PBHs are constrained by the products of the early stages of their evaporation (Sec.~\ref{sec:evap}), planetary and stellar mass and heavier PBHs are constrained by gravitational lensing observations (Sec.~\ref{sec:lensing}), with stellar mass PBHs also being constrained by their dynamical effects (Sec.~\ref{sec:dynamical}), the consequences of accretion onto them (Sec.~\ref{sec:accretionconstraint}) and gravitational waves from their mergers (Sec.~\ref{sec:gwmergers}). The abundance of PBHs formed by the collapse of large density perturbations is also constrained indirectly via constraints on the amplitude of the power spectrum (Sec.~\ref{sec:indirect}). In the past few years there has been significant activity on PBH abundance constraints, with new constraints being proposed and existing constraints being revisited. In some cases (e.g.~interactions with stars, Sec.~\ref{sec:stars}, or GRB femtolensing, Sec.~\ref{femto}) `old' constraints have been shown not to hold. While individual constraints may have significant modelling uncertainties the Solar mass region is now subject to multiple constraints. If the constraints are taken at face value, PBHs with masses in the planetary to multi-Solar mass range can only make up a subdominant fraction of the DM. However the robustness of this conclusion depends on the late-time clustering of the PBH population, which remains unclear. 
The asteroid mass region ($10^{17} \, {\rm g} \lesssim M_{\rm PBH} \lesssim  10^{22} \, {\rm g}$) remains open. This largely reflects the difficulty of detecting such light compact objects.

Primordial Black Holes, in particular asteroid mass ones, remain a viable dark matter candidate.
Further improvements in the theoretical calculations of the production and evolution of PBHs are required to reliably predict the abundance and properties of PBHs from a given model.
Even so, it seems clear that a cosmologically interesting number of PBHs can only be produced in specific models of the early Universe, and often fine-tuning is required. 
However, whether or not PBHs are a significant component of the DM is a question that has to be answered observationally. Novel ideas are needed here to either detect or rule out the remaining open parameter space.

\section*{Acknowledgments}

AMG is supported by STFC  grant ST/P000703/1. BJK thanks the Spanish Agencia Estatal de Investigaci\'on (AEI, MICIU) for the support to the Unidad de Excelencia Mar\'ia de Maeztu Instituto de F\'isica de Cantabria, ref. MDM-2017-0765. 
We acknowledge
  the use of {\sc NumPy} \citep{vanderWalt:2011bqk} and  {\sc Matplotlib} \citep{Hunter:2007ouj}. 
\href{https://automeris.io/WebPlotDigitizer}{\underline{WebPlotDigitizer}} has been used to extract data from publications. 
We are grateful to Chris Byrnes, Bernard Carr, Karim Malik, Jordi Miralda-Escude and Teruaki Suyama for useful discussions and/or comments.
BJK thanks Adam Coogan, Zu-Cheng Chen and Mohsen Ghazi for contributions to the PBHbounds code.

\bibliographystyle{apsrev4-1}
\bibliography{PBHDMbibfile}

\end{document}